# Design studies of a continuous-wave normal conducting buncher for European X-FEL

Shankar Lal[1,−], V. Paramonov[2], H. Qian[1,=], H. Shaker[1], G. Shu[1], Ye Chen[1], F. Stephan[1]

[1]Deutsches Electron Synchrotron DESY, Plataneallee 6, 15738 Zeuthen, Germany
[2]Institute for NuclearResearch of RussianAcademy of Sciences, 60-th OctoberAnniversary Prospect 7A, 117312, Moscow, Russia

*Abstract*

A three-cell 1.3 GHz, Normal Conducting (NC) buncher is designed for a possible future upgrade of the European XFEL to operate in a continuous-wave (CW) / long pulse (LP) mode. The RF geometry of the buncher is optimized for high shunt impedance, large mode separation as well as multipacting free in the operation range. The bunchersupport cavity voltage of 400 kV with an RF power dissipation of 14 kW. A tapered waveguide-based RF power coupler is designed to feed the RF power to the buncher. The RF power coupler port is optimized for field asymmetry compensation. The thermal load due to RF power dissipation is analyzed using Multiphysics simulations in CST and a simplified cooling scheme is designed.

## 1. Introduction

The European Xray Free Electron Laser (Eu-XFEL) is operating in pulsed mode with a duty factor of ~ 0.65%, and the injector system is based upon a 1.6 cell, L-band normal conducting (NC) pulsed photocathode RF gun [1,2]. For possible future upgrade to operate the E-XFEL in a continuous wave (CW) or in a Long Pulse (LP) mode it requires a CW injector system. There are two types of CW electron guns under development for Eu-XFEL, the main option is the superconducting (SC) gun [3], and the backup option is a NC, Very High Frequency (VHF) gun similar to the APEX gun at LBNL [4, 5].

The normal conducting VHF gun-based injector system is being studied at Photo Injector Test Facility at DESY Zeuthen site (PITZ) [5]. The proposed injector system comprises a 217 MHz VHF gun, a 1.3 GHz buncher, two solenoids and a SC linac module [4,5,6]. A schematic diagram of the proposed injector system is shown in Fig.1.

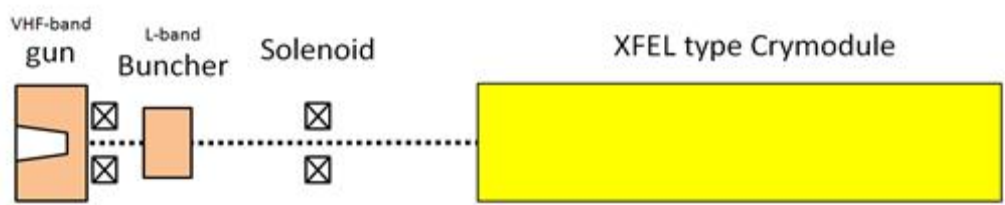

Fig. 1: Schematic diagram of a VHF gun-based CW injector.

---

[−]Presently at Raja Ramanna Centre for Advanced Technology (RRCAT), Indore, MP, India.
Email: shankar.lal@desy.de, shankar@rrcat.gov.in
[=] houjun.qian@desy.de

The gun frequency is the 6[th] sub-harmonic of 1.3 GHz (frequency of the SC linac), so every RF bucket of the gun synchronizes with the linac. Due to the relatively low energy (<1 MeV) and low RF frequency, the 'cigar' mode photoemission, i.e. long laser duration, can be used to minimize the beam emittance by reducing emission peak current. The 1.3 GHz buncher increases the beam peak current in the low energy drift before the beam enters the SC linac. Following the APEX experience, we considered a 1.3 GHz buncher with a cavity voltage of ~400 kV predicted by beam dynamics studies [6].

The RF and power coupler design are presented in section 2. The effect of RF power coupling slot on electromagnetic field and beam parameters, and optimization of coupler cell are discussed in section 3, which is confirmed by beam dynamics simulations in section 4. The multipacting and Multiphysics simulations are presented in section 5 and 6 respectively. An evaluation of geometrical tolerances for the resonance frequency and its tuning method is presented in section 7, followed by conclusion in section 8.

## 2. Design of the buncher

### 2.1. Literature survey

The normal conducting buncher cavities are employed in the injector system for many Energy Recovery Linac (ERL) and Free Electron Laser (FEL) projects [9-12]. The Cornell/JLab ERL injector system employs a 1.3 GHz, single cell re-entrant type buncher cavity. The 120 kV buncher compresses the electron bunches of $\sigma_t$ =12 ps from a 400 kV DC photocathode gun, to $\sigma_t$ =2.3 ps before injecting into the superconducting linac [13]. The compact ERL (cERL) at KEK uses a 1.3 GHz, single cell re-entrant type buncher cavity. The 130 kV buncher compresses the 3-10 ps bunches from a 500 kV DC photocathode RF gun to <1-3 ps [14]. The injector system of the Advanced Photo-injector Experiment (APEX) at LBNL, uses a 1.3 GHz two-cell re-entrant type buncher, to compress the ~60 ps bunches from a VHF gun to ~10 ps with a cavity voltage of 240 kV [15]. The important RF parameters of CW buncher cavities discussed above are summarized in Table 1. In this paper the cavity voltage and the shunt impedance refers to the effective voltage $V_e = \left| \int_0^L E_z(z) e^{-i\frac{2\pi z}{\beta \lambda}} dz \right|$ and shunt impedance $Z_e = \frac{V_e^2}{P_c}$, where L is the length of the RF accelerating structure, $\beta = v/c$ is the normalized velocity of the charge particle with the velocity of light, λ is the wavelength of the RF field and $P_c$ is the power dissipation in the cavity.

Table 1: Comparison of CW buncher cavities used in different ERL/FEL projects.

| Parameters | Cornell/Jlab | cERL (KEK) | APEX (LBNL) |
|---|---|---|---|
| No. of cells | 1 | 1 | 2 |
| Operating frequency (MHz) | 1300 | 1300 | 1300 |
| $Z_e$(MΩ) | 4.2 | 5.33 | 7.8 |
| Nominal $V_e$(kV) | 120 | 130 | 240 |
| Power dissipation (kW) | 3.5 | 3.17 | 7.4 |
| Peak surface field (MV/m) | 8.8 | 4.8 | 4.7 |
| Kilpactric limit (MV/m) | 32 | 32 | 32 |
| Power density (W/cm$^2$) | / | 6.9 | 5.8 |
| Power dissipation (kW) for $V_e$ = 400 kV | 38 | 30 | 21 |

## 2.2. RF design of two-cell buncher

Based on the summary in Table 1, for the desired accelerating voltage of 400 kV the RF power dissipation of APEX 2-cell buncher is the lowest around 21 kW. The PITZ gun cooling can remove 40 kW to 60 kW from a 2-cell L-band cavity [16,17], so in principle the APEX type buncher should work with enhanced water-cooling designs. In order to further simplify the buncher cooling, new designs of higher shunt impedance are studied to reduce the RF heating.

Several two-cell buncher cavities are simulated using 3D electromagnetic code CST Microwave Studio® (CST MWS) [18]. The first buncher design with cell geometry similar to the KEK design is shown in Fig. 2, and the major RF parameters are summarized in Table 2. This design has RF power dissipation of 17 kW as compared to 21 kW of the APEX buncher for the desired accelerating voltage of 400 kV. Though the RF power dissipation is reduced, it has two major limitations: (1) the mode separation between the '0' and 'π' mode is ~1 MHz, which means each cell has to be coupled to RF power independently to avoid mode mixing, just like the APEX buncher, (2) the maximum heating is near the inter-cell coupling iris which is difficult to remove due to space constraints.

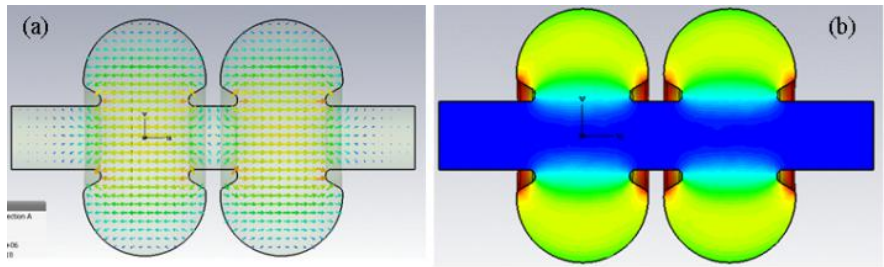

Fig. 2: Two-cell buncher (a) electric field array plot and (b) magnetic energy density distribution for π mode predicted by CST MWS.

To increase the mode separation the inter-cell coupling iris is modified from a nose-cone shape to elliptical like a TESLA cavity as shown in Fig. 3. While important RF parameters are given in Table 2. The modified structure has a mode separation of ~3 MHz, but the RF power reduction compared to APEX buncher design is not significant, around 10%.

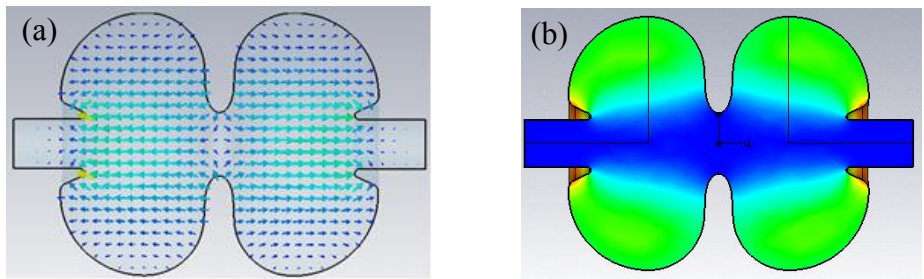

Fig. 3: DESY two-cell buncher (a) electric field array plot and (b) magnetic energy density distribution for π mode predicted by CST MWS.

Table 2: Comparison of RF parameters two-cell buncher of different geometries.

| RF parameters | KEK type | DESY design |
|---|---|---|
| $f_\pi$ (GHz) | 1.3 | 1.3 |
| Quality factor $Q_0$ | 25316 | 27819 |
| $Z_e$ (MΩ) | 9.9 | 9.19 |
| Mode separation $f_\pi - f_0$ (MHz) | 1.03 | 3.02 |
| $P_c$ (kW) for 400 kV | 17 | 18 |

**2.3. RF design of three-cell buncher**

To further reduce the RF power dissipation, a TESLA like elliptical cell is added into the two-cell design to form a three cell buncher, as shown in Fig.4. The geometrical dimensions are re-optimized to achieve the π mode at 1.3 GHz with an equal on-axis field amplitude in all the cells. Since a three-cell coupled structure supports three Eigen modes viz. '0', 'π/2' and 'π'. The on-axis electric fields for different Eigen modes along the structure length are shown in Fig. 5. Since the number of modes are increased, the separation of π mode from the nearest mode (π/2) is reduced for the same inter-cell coupling iris. The inter-cell coupling iris diameter is enlarged from 42 mm to 56 mm, and the mode separation between the π and the π/2 mode is 3.18 MHz. The RF power dissipation for 400 kV is ~13 kW and RF heating per cell is lowered to ~4.3 kW, more than a factor of 2 lower than the ~9 kW/cell of the two-cell case. The RF parameters of the three cell buncher are summarized in Table 3. The 3-cell buncher design has a similar heating load per cell as the APEX buncher for the cavity voltage of 240 kV, i.e. ~4 kW/cell, and its cooling challenge is expected to be reduced compared to the other designs, so it is selected for further studies.

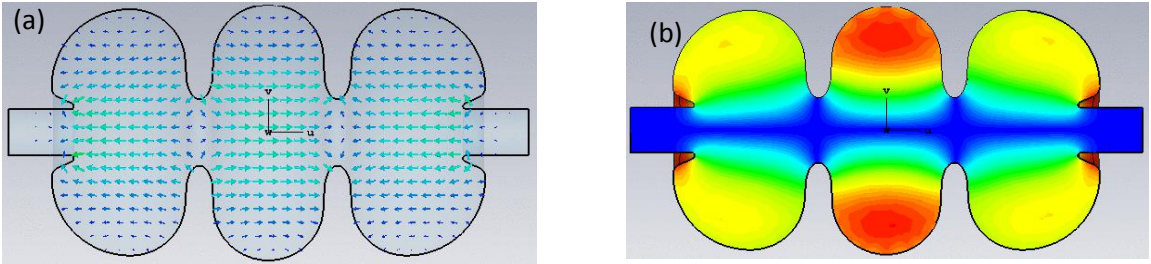

Fig. 4: (a) Electric field array plot and (b) magnetic energy density distribution for π mode predicted by CST MWS.

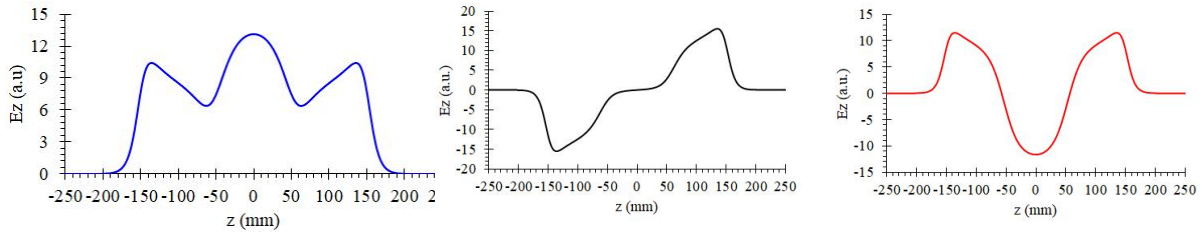

Fig.5: on-axis field profile for (a) 0 mode, (b) π/2 mode and (c) π mode.

Table 3: RF parameters of the three cell buncher.

| Parameters | Values |
|---|---|
| Resonance frequency $f_\pi$ (GHz) | 1.3 |
| Quality Factor $Q_0$ | 27639 |
| Ze(MΩ) | 12 |
| Mode Separation $f_\pi$-$f_{\pi/2}$ (MHz) | 3.18 |
| Power dissipation Pc (kW) for 400 kV | 13 |

### 2.4. RF power coupler design

In principle RF power can be coupled to an RF cavity by using a waveguide or a coaxial loop or antenna type coupler [19]. Each type of coupler has its pros and cons, and a coupler is selected depending upon operating frequency and power. A coaxial (loop and antenna) type coupler is compact and easy to tune the coupling; however, it has lower power handling capacity and difficult to cool. On the other hand, waveguide-based coupler is bulky but it has larger power handling capacity and easy to cool [19]. The coaxial couplers are more vulnerable to multipacting as compared to the waveguide-based couplers. Finally, the availability of commercial RF windows is also important for selecting a coupler. Considering the operating frequency of 1.3 GHz, and the RF power dissipation of 13 kW a WR650 rectangular waveguide (dimension:165.1 mm x 82.55 mm) based RF power coupler is chosen for the buncher. Since the buncher is a standing wave multi-cell accelerating structure, in-principle RF power can be fed to any of the cell, however RF power feeding in the middle cell eliminates the possibility of excitation of the 'π/2' mode, as there is no field in the middle cell for this mode (see Fig.5). Since there is no excitation of the 'π/2' mode, the nearest mode to the π mode is the 0 mode in this case and the practical mode separation is increased from ~3 MHz to ~6 MHz. Considering this advantage, the RF power coupler is designed on the middle cell (this cell will be called as coupler cell in further text). The RF power is coupled from the waveguide to the coupler cell by a slot on its outer wall. The ratio of the RF power coupled to the cavity to the input RF power is defined in terms of the RF power coupling coefficient ($\beta_{RF}$) as

$$\beta_{RF} = \frac{P_{ext}}{P_c}, \qquad (1)$$

where $P_c$ is the power dissipation in the cavity and $P_{ext}$ is the power radiated from cavity towards the waveguide when RF power is turned off [19-20]. For a waveguide to a single cell where RF power is coupled to the cavity by an elliptical hole at the interface of waveguide and cavity, the $\beta_{RF}$ is given as [20]

$$\beta_{RF} = \frac{\pi^2 Z_0 k_0 \Gamma_{10} e_0^4 l_1^6 e^{-2\alpha d}}{9\, ab(K(e_0)-E(e_0))^2} \frac{H_1^2}{P_c}, \qquad (2)$$

where $H_1$ is the magnetic field at location of slot when there is no slot, $Z_0$ =377 Ω is the impedance of free space, $k_0=2\pi/\lambda$ is the wave number, $l_1$ is the major-semi axis of the ellipse equivalent to RF coupling slot, $e_0 = \sqrt{1-(l_2/l_1)^2}$, $l_2$ is the minor-semi axis of the ellipse equivalent, α is the attenuation, d is the thickness of the RF coupling slot, $a$ and $b$ are the wider and shorter dimensions of the waveguide, $\Gamma_{10} = k_0\sqrt{1-(\lambda/2a)^2}$, and K($e_0$) and E($e_0$) are the complete elliptic integrals of the first and second kind [20]. For an input RF power ($P_{in}$), the power coupled to the cavity is given as

$$P_c = \frac{4\beta_{RF}}{(1+\beta_{RF})^2} P_{in}. \tag{3}$$

The power coupling to the cavity is maximum for $\beta_{RF}$ =1 (known as critical coupling).

Since the presence of an RF coupling slot reduces the resonance frequency of coupler cell [21], so coupling tuning simulations are not straightforward for multi-cell cavity as compared to single cell cavity. Therefore, while optimizing the dimensions of the RF coupling slot to achieve the desired value of $\beta_{RF}$, its effect on the coupler cell resonance frequency also needs to be compensated to simultaneously tune both the π mode frequency as well as the field flatness (defined as the ratio of on-axis electric field amplitude in each cell to that in the coupler cell). As discussed in Ref [21], for a coupled multi-cell structure, the resonance frequency, field flatness and $\beta_{RF}$ are interdependent which makes the tuning very tedious and time consuming. Although, there is an optimizer available in the CST to automatically optimize the multiple geometrical dimensions (RF coupling slot and ID's of the cells) to achieve multiple RF parameters, however it also very time consuming as simulations are performed multiple times.

To simplify the tuning procedure, we adopted the approach discussed in Ref [21-22] in which the coupler slot tuning is decoupled from tuning of the frequency and the field flatness for the multi-cell cavity. First, we determined the ratio of $H_1^2/P_c$ for the full structure tuned for $f_\pi$ =1.3 GHz with uniform field flatness. Then the resonance frequency of the coupler cell is determined by de-tuning end-cells by increasing their IDs such that there is no field in them, and also determine the radio of $H_1^2/P_c$. An RF model of the buncher to determine the independent coupler cell parameters with on-axis electric field is shown in Fig.6. After determining the independent coupler cell parameters, the value of the independent coupler cell to the waveguide coupling coefficient ($\beta_{RF,c}$) can be predicted by following formula

$$\beta_{RF,c} = \frac{(H_1^2/P_c)_c}{(H_1^2/P_c)_\pi} \beta_\pi, \tag{4}$$

here c stands for coupler cell. The values of $H_1$, $P_c$, and coupling coefficient for π mode and for the independent coupler cell are given in Table 4. For our case to achieve $\beta_\pi$=1 the coupler cell need to be tuned for $f_{rc}$=1296.925 MHz with $\beta_{RF,c}$=2.17 to achieve the π mode with uniform field flatness. The major advantage of this approach is that the tuning of coupling coefficient is decoupled from field flatness and at a time only one cell needs to be tuned. Although as discussed above the presence of the RF coupling slot reduces the resonance frequency of the coupler cell which needs to be compensated by varying its ID. The required variation in the ID of coupler cell either can be estimated analytically as discussed in Ref [21, 22] or frequency can be directly tuned using electromagnetic solver.

Table 4: RF parameters of the buncher and independent coupler cell.

| Parameters | Frequency (MHz) | $H_1$(A/m) x $10^4$ | $P_c$ (W) x$10^5$ |
|---|---|---|---|
| π mode | 1300 | 1.862 | 2.961 |
| Coupler cell | 1296.925 | 2.744 | 2.967 |

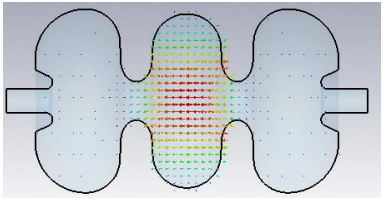 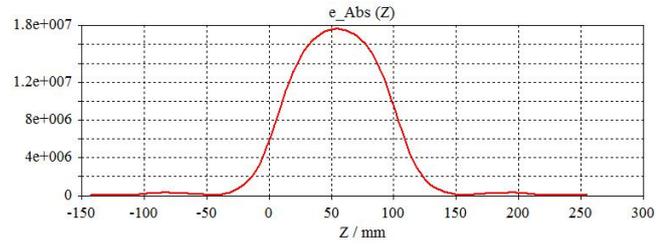

Fig. 6: RF model and on-axis electric field profile of three cell buncher to determine the coupler cell parameters.

As discussed in Ref [23], the presence of an RF coupling slot also breaks the azimuthal symmetry of the coupler cell which leads to generation of multipole modes and may distort the beam trajectory, shape and phase space. The strength of multipole modes increases with the dimensions of the RF coupling slot, therefore the shape of the RF coupling slot needs to be chosen such that the desired $\beta_{RF}$ can be achieved with minimal size of the slot. An oblong slot shape was chosen as it is widely used for RF coupling. For an oblong slot, the value of $\beta_{RF}$ mainly depends upon its length (see Eqn. 2) and its width need to be chosen to avoid arcing. As many S-band RF accelerating structure around the world uses oblong slot to feed RF power, having slot width of 9.5 mm, hence we scaled it and chosen slot width of 20 mm. After fixing the width, the length of the RF coupling slot predicted using Eqn. 2 is 69 mm, which is very long and may have very high perturbation in electromagnetic field in the coupler cell. The length of the slot can be reduced by tapering the waveguide in shorter dimensionto increase the magnetic field near the slot [24]. The value of $\beta_{RF}$ in such a case depends upon the tapering angle. Considering the machining constrains, the width of the waveguide at cavity interface is fixed to 30 mm (5 mm margin on both sides of the coupler slot) while the non-tapered side is fixed to the WR650 dimensions (165.1mm x82.55 mm). To optimize the tapering angle, variation in $\beta_{RF,c}$ versus tapering length is studied For a fixed RF coupling slot length, the value of $\beta_{RF,c}$ is maximized by varying the tapered waveguide length, then length of RF coupling slot is varied to achieve the desired $\beta_{RF,c}$. The desired value of $\beta_{RF,c}$ is achieved for an RF coupling slot length of 41 mm while waveguide length of 130 mm which corresponds to a tapering angle of ~11.5°.The variation in $\beta_{RF,c}$ with the length of the waveguide is shown in Fig. 7.

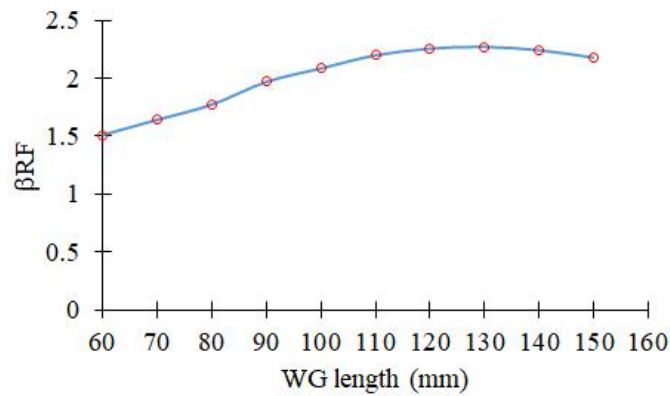

Fig. 7: Variation in $\beta_{RF}$, with the length of the waveguide with shorter dimension at the cavity interface of 30 mm and 82.55 mm at another end, for coupling slot dimensions of 41mm x 20 mm, for π mode

Due to RF coupling slot, resonance frequency of the coupler cell ($f_{rc}$) is reduced from 1296.925 MHz to 1294.89 MHz. To compensate the reduction in $f_{rc}$, the ID of the coupler cell is reduced to 99.77 mm from 99.91 mm. With the ID of the coupler cell and the dimensions of the RF coupling slot, the full structure resonates for π mode at 1.3 GHz with $\beta_{RF}=1.05$ and field uniformity of 1.01 (ratio of the on-axis electric field amplitude in end cell to that in the coupler cell). The shunt impedance of the buncher is reduced to 11.63 MΩ from 12 MΩ due to presence of the coupler slot. A view of the buncher with the RF coupler details is shown in Fig. 8, while the RF parameters are given in Table 5.

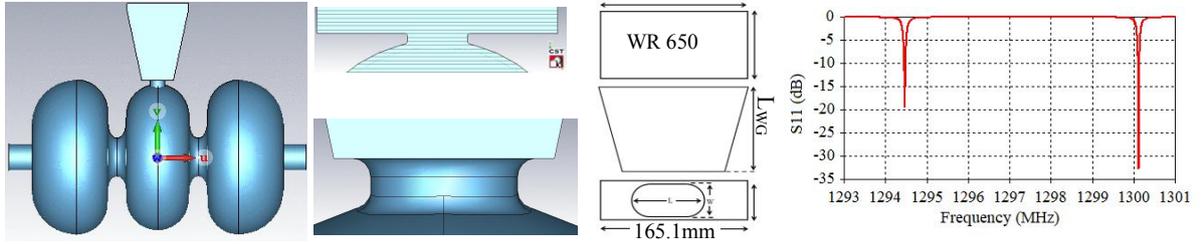

Fig.8: CST model of the three-cell buncher with waveguide and RF coupling slot details. The frequency spectrum predicted by frequency domain solver of CST is also shown.

Table 5: RF parameters of the buncher with power coupler

| | |
|---|---|
| $f_\pi$ | 1300 MHz |
| $Q_0$ | 27060 |
| $Z_e$ (830 keV beam) | 11.63 MΩ |
| $S_{11}$ | -32(dB) |
| $\beta_{RF}$ | 1.05 |
| RF slot size (L×W) | 41mm × 20 mm |

## 3. Effect of RF coupling on field symmetry and optimization of coupler cell

As discussed in the previous section, presence of the RF coupling slot breaks azimuthal symmetry of the coupler cell, leading to multipole modes. Hence, the electromagnetic field in the coupler cell can be decomposed into a sum of multipole angular modes [23,25]. As discussed in Ref [23,25] the longitudinal electric field in such a case can be written as

$$E_z^{TM10}(r,\theta,z) = E_0\cos(kz)\sum_{m=0}^{\infty} a_m J_m(k_{cm}r)\cos m(\theta - \theta_0), \qquad (5)$$

where 'm' is the azimuthal index and describes the number of periods of the field in 360° rotation of azimuthal angle, $a_m$ is the normalized amplitude of the $m^{th}$ multipole mode with respect to the monopole mode (m=0), $k_m$ is the root of the Bessel function of order 'm' with $J_m(k_{cm}R) = 0$, where R is the cell (cavity) radius, 'r' and 'θ' are the cylindrical coordinates and $\theta_0$ is the polarization angle. The mode with index m=0 is known as a monopole mode, and it is azimuthally symmetric. The modes with index m =1, 2, 3 and 4 are known as the dipole mode, the quadrupole mode, the sextupole mode and the octupole respectively. Due to presence of the multipole modes the electromagnetic field in the coupler cell gets distorted which causes a distortion in the beam shape and leads to emittance growth. For no space charge cases, with assumption that β=1,the contributions of different multipole modes to the normalized rms emittance of the beam are given by [23]

$$\varepsilon_{nx}^{010} = \frac{k_0^2 \alpha L cos\overline{\varphi}_0 \sigma_x^2 \sigma_\varphi^2}{2}, \varepsilon_{ny}^{010} = \frac{k_0^2 \alpha L cos\overline{\varphi}_0 \sigma_y^2 \sigma_\varphi^2}{2} \qquad (6)$$

$$\varepsilon_{nx}^{110} = 0, \varepsilon_{ny}^{110} = \frac{a_1 k_1 \alpha L cos\overline{\varphi}_0 \sigma_y \sigma_\varphi}{2} \qquad (7)$$

$$\varepsilon_{nx}^{210} = \frac{a_2 k_2^2 \alpha L cos\overline{\varphi}_0 \sigma_x^2 \sigma_\varphi}{4}, \varepsilon_{ny}^{210} = \frac{a_2 k_2^2 \alpha L cos\overline{\varphi}_0 \sigma_y^2 \sigma_\varphi}{4} \qquad (8)$$

$$\varepsilon_{nx}^{310} = \frac{a_3 k_3^3 \alpha L \sigma_x^3 \sqrt{sin^2\overline{\varphi}_0 + cos\overline{\varphi}_0 \sigma_\varphi^2}}{8}, \quad \varepsilon_{ny}^{310} = \frac{a_3 k_3^3 \alpha L \sigma_y^3 \sqrt{sin^2\overline{\varphi}_0 + cos\overline{\varphi}_0 \sigma_\varphi^2}}{8} \qquad (9)$$

$$\varepsilon_{nx}^{410} = \frac{\sqrt{6} a_4 k_4^4 \alpha L \sigma_x^4 \sqrt{sin^2\overline{\varphi}_0 + cos\overline{\varphi}_0 \sigma_\varphi^2}}{96}, \quad \varepsilon_{ny}^{410} = \frac{\sqrt{6} a_4 k_4^4 \alpha L \sigma_y^4 \sqrt{sin^2\overline{\varphi}_0 + cos\overline{\varphi}_0 \sigma_\varphi^2}}{96}, \qquad (10)$$

where $\varepsilon_n$ is the normalized rms emittance, $\sigma_{x,y}$ is the transverse beam size in x, y direction and $\sigma_\varphi$ is the longitudinal rms beam size in terms of the rf phase ($\sigma_\phi=\omega\sigma_t$),$\overline{\varphi}_0$ is the mean phase of the bunch in the cavity ($\varphi_0$=0 means max acceleration), $\alpha = (eE_0/2mc^2 k)$ is the normalized RF field strength in the cavity. Here, it is assumed that the strength of the multipoles is very weak as compared to the monopole mode such that there is negligible correlation between x, y and $\varphi_0$, and the beam remains symmetric in the buncher ($<x^2> = <y^2>$, $<x^4> = <y^4>$).

As described in detail in Ref [25], the relative amplitude of the multipole modes can be determined by analyzing the azimuthal variation of Ez, which can be found using a 3D electromagnetic code like CSTMWS in our case. The electric field at a longitudinal location (say, $z_0$) and at a radius (say '$r_o$') can be written as a Fourier series

$$E_z(\theta) = E_0 \, coskz_0 \left(1 + \sum_{m=1}^{4} A_m cosm(\theta - \theta_0)\right) \qquad (11)$$

Here, $A_m$ is the normalized amplitude of the m[th] multipole mode, which can be calculated from Fast Fourier Transformation (FFT) of $E_z(\theta)$ predicted by Eigen mode solver of CST MWS. The amplitudes of multipole mode ($a_m$) used in emittance calculation are related to those determined by CST MWS as

$$a_m = \frac{A_m J_0(k_{c0} r_0)}{J_m(k_{cm} r_0)} . \qquad (12)$$

Once the normalized amplitude of the multipole modes is known, their contributions to the normalized rms emittance can be predicted using Eqn. (6-10).

### 3.1. Determination of the amplitude of Multipole modes

As discussed above the normalized amplitude of multipole modes can be calculated by analyzing the variation of Ez (θ) at the centre of coupler cell, which can be determined by using postprocessor of the Eigen mode solver of the CST MWS [18]. Since for Eigen mode/ Frequency domain solver it is important to optimize the meshing to construct the outer geometry only while the mesh in inner volume of the cavity is not very important for the purpose of resonant frequency calculation, hence CST MWS automatically makes finer mesh near the boundary and coarse meshing in the central region. Therefore, variation of Ez with θ cannot be determined accurately. In-order to achieve good accuracy with reasonable simulation time, a local fine mesh is used near the cavity axis, as shown in Fig.9.

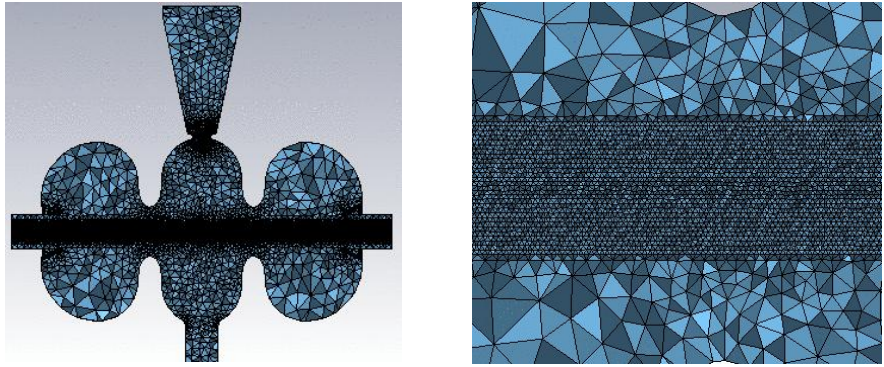

Fig.9: Meshing view of the buncher showing local mesh refinement near axis.

Since the effect of a port opining on the RF field is maximum near the port and minimum around the axis. In order to take a worst-case scenario, the Ez (θ) is analyzed at $r_0 = 4\sigma_r$, =10 mm [6], where $\sigma_r$ is the beam rms size in the buncher. The variation of $E_z(\theta)$ at longitudinal center of the coupler cell) and at the radius of 10 mm is shown in Fig. 10. From Fig. 10, it is clear that the presence of the RF power coupler slot excites a dipole mode. A comparison of normalized amplitude of different multipole modes calculated by FFT of Ez is given in Table 6. For the electron beam with $\sigma_{x,y}$= 2.48 mm, $\sigma_z$= 3.68 mm, and $E_o$ =2.4 MV/m (cavity voltage of 400 kV) with an RF phase of -35.8°, the expected emittance due to dipole is ~0.044 *mm mrad* which is significantly high considering the final desired beam emittance of the order of 0.1 *mm mrad*. The emittance growth due to quadrupole mode is 4.9e-4 *mm mrad*. Although the pure RF quadrupole effect seems to be small, but it might get amplified by space charge effect, or cause transverse phase space coupling after Larmor rotation in the solenoid. So, both RF dipole and quadrupole distortion needs to be minimized.

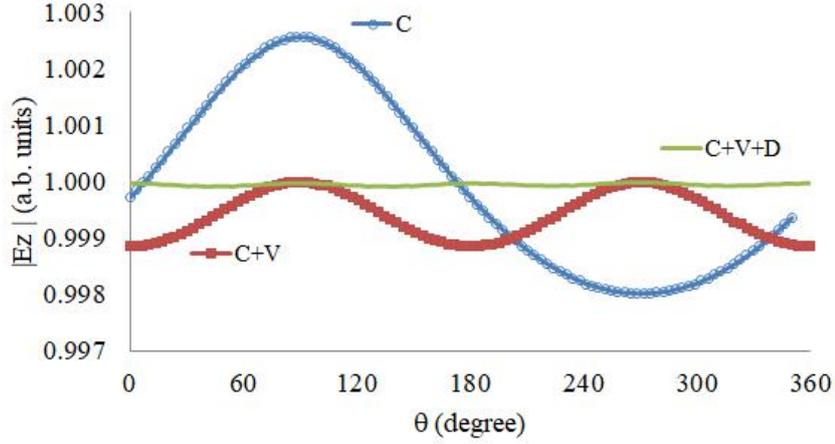

Fig. 10: Variation of normalized Ez with θ at centre of coupler cell and at $r_0$ =10 mm for different ports configurations. Here C is stand for the coupler port, C+V is for the coupler and the vacuum port and C+V+D is for the coupler, vacuum port and the dummy ports are present.

Table 6: Normalized amplitude of the multipole modes for different configurations of the coupler cell of the buncher.

| Configuration | $A_1$(dipole) | $A_2$(quadrupole) | $A_3$(sextupole) | $A_4$(octupole) |
|---|---|---|---|---|
| RF coupler port | 1.11e-3 | 1.43e-4 | 3.04e-5 | 6.53e-6 |
| RF coupler + vacuum port | 6.91e-6 | 2.87e-4 | 4.74e-6 | 9.56e-6 |
| RF coupler + vacuum + dummy ports | 4.57e-6 | 2.23e-6 | 7.35e-7 | 1.49e-5 |

### 3.2. Optimization of coupler cell

As discussed in Ref [23], the dipole mode can be reduced by adding an additional port diametrically opposite to the RF coupler port, which can also be served as a port for vacuum pumping (and known as vacuum port). The vacuum port must reduce the dipole mode but should also prevent or minimize the leakage RF power. Generally, the vacuum port has a shape similar to the RF coupling slot with a cylindrical pipe [23, 24]. As discussed in previous section, the RF coupling slot is of oblong shape, hence generally the vacuum port is also an oblong slot [23,24]. To reduce the dipole mode as well as the field enhancement, we made the vacuum hole of a circular shape, which reduced its dimension along the azimuthal direction by ~20% compared to the RF coupling hole when dipole component is compensated. To further reduce the field enhancement, the edge between the vacuum hole and the body of coupler cell is rounded with a radius of 8 mm. A vacuum hole of circular shape also has an advantage of higher vacuum pumping conductance as compared to the oblong slot. The optimum vacuum port (hole) diameter is 34.6 mm to minimize the dipole mode. With the optimized vacuum port, the variation in Ez with θ is shown in Fig. 10, and the normalized amplitudes of different multipoles are given in Table 6. The presence of vacuum port reduces the π mode frequency of coupler cell by 2.1 MHz which is compensated by reducing its ID to

99.62 mm from 99.77 mm. The shunt impedance is also reduced slightly to 11.50 MΩ from 11.63 MΩ.

Adding vacuum port, reduced the dipole mode to 6.91e-6 from 1.11e-3, but it increased the quadrupole mode to 2.87e-4 from 1.43e-4. As discussed in Ref [23, 25], the quadrupole mode can be reduced by adding two ports orthogonal to the RF/vacuum ports. These ports are known as quadrupole mode compensation dummy ports. As discussed in Ref [23,25], generally the dummy ports also have slots of shape similar to the RF port. Again, to reduce the field enhancement, dummy ports are also modelled as circular shape. Generally, the dummy port has through hole which is blanked using a CF flange to achieve the vacuum sealing. To minimize the brazing joints, we modelled dummy ports as blind holes. The edge of the dummy holes to cavity is also rounded with radius of 8 mm, to minimize the field enhancement. Figure 11 shows the model of the buncher with all the ports. The optimum IDs of the dummy holes for quadrupole mode compensation depends upon the length of the dummy holes. A variation in the optimum ID of the dummy holes (ports) with length is shown in Fig.12. The ID of the dummy port reduces with length and saturates after length of 20 mm.

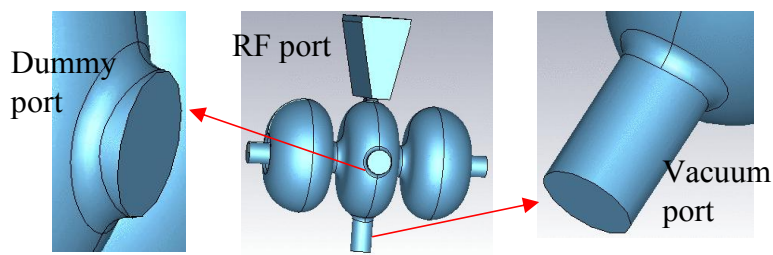

Fig.11: Model of the buncher with the RF, vacuum and the dummy ports.

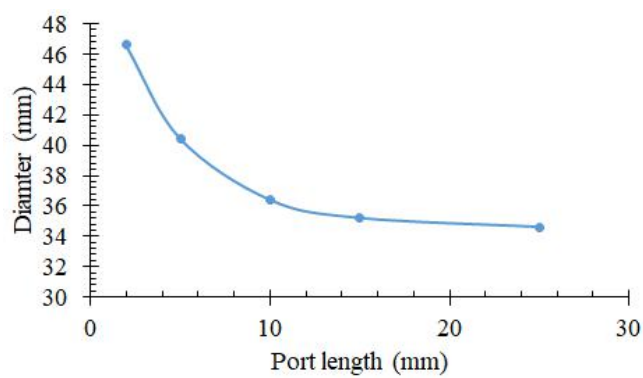

Fig. 12: Variation in diameter of the dummy ports with length for suppression of quadrupole mode.

Considering the strength of the material at the location of the dummy port and machinability, the dummy holes of 5 mm length are considered. The optimum size of the dummy holes is 40.45 mm. With all ports, variation of Ez with θ is shown in Fig.10, and the normalized amplitudes of different multipoles are given in Table 6. By adding dummy holes the amplitude of quadrupole mode is reduced by two orders of magnitude, however the octupole mode is increased to twice. The enhancement of octupole mode can be ignored as the emittance growth due to octupole mode is negligible. The reduction in resonance frequency due to dummy holes is compensated by reducing the coupler cell ID to 99.35 mm. The variation in other RF parameters viz. quality factor and shunt impedance is negligible. The expected emittance growth due to multipole modes for different configurations is given in Table 7.

Table 7: Contribution to the normalized emittance (in mm.mrad) due to different multipole modes for different configurations of the coupler cell of the three cell buncher (no space charge estimation).

| Configuration | $\varepsilon_y^{110}$ (Dipole) | $\varepsilon_y^{210}$ (Quadrupole) | $\varepsilon_y^{310}$ (Sextupole) | $\varepsilon_y^{410}$ (Octopole) |
|---|---|---|---|---|
| Coupler port | 0.044 | 4.9e-4 | 9.08e-5 | 1.38e-6 |
| Coupler +vacuum port | 2.74e-4 | 9.92e-4 | 1.42e-6 | 2.02e-6 |
| Coupler + vacuum+ 2 dummy ports | 1.82e-4 | 7.70e-6 | 2.20e-6 | 3.16e-6 |

### 3.3. Evolution of emittance growth induced by phase asymmetry

Although the emittance growth due to multipole modes with a coupler port, a vacuum port and two dummy ports in the coupler cell is two orders of magnitude lower as compared to the desired value. However due to RF loss in the buncher walls, there will be a travelling wave from the RF coupling ports towards vacuum port [26] which gives a phase asymmetry along the direction of the power flow, and may increase the emittance of the beam. The emittance growth due to phase asymmetry is given by [26]

$$\varepsilon_{ny}^{phase} = \frac{1}{2}\alpha k_y L \sigma_y \sigma_\varphi \sin\overline{\varphi}_0 , \qquad (13)$$

where $k_y$ is the wave number of the travelling wave. Since Eigen mode solver of CSTMWS calculates standing wave only, variation in the phase along power flow direction is predicted by the frequency domain solver of CSTMWS. The variation in phase along y direction (direction in which RF coupler is located, see Fig.8) is shown in Fig.13. The value of $k_y$ is 0.00338 rad/m, and the estimated emittance growth is 0.0015 mm mrad, which is negligible.

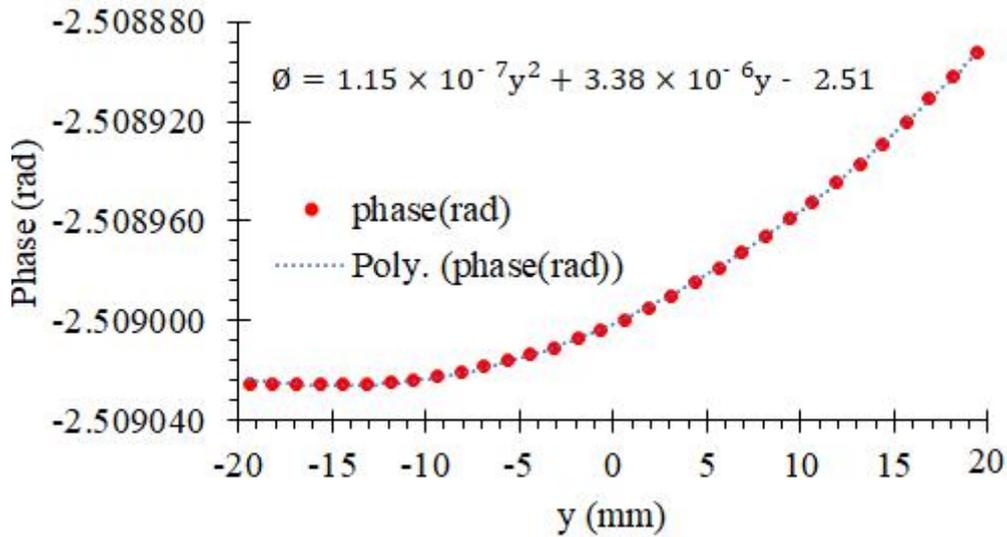

Fig. 13: Phase of Ezvs y direction in the longitudinal symmetry plane of the coupler cell with the coupler + vacuum +dummy ports.

### 3.4. RF pickup design

A pickup loop located in the vacuum pipe is designed to sample the RF field in the coupler-cell . The pickup signal will be used for feedback, so the dimensions of the pickup loop are optimized to have RF signal of > 0.5 W, required for processing electronics. A view of the pickup loop is shown in Fig.14. Since the pickup power is very low, the inner and outer conductor dimensions are chosen as of the N type connector. The center of the pickup loop is located 30 mm away from the inner surface of the buncher. The strength of the pickup signal is predicted by S21 using Frequency domain solver of the CSTMWS. A typical plot of $S_{21}$ from coupler port to pick-up is shown in Fig.15. A variation in the pickup signal with length of the loop is shown in Fig.16. In presence of pickup loop, the multipole analysis shows that pickup loop has almost no effect on multipole modes. Since the pickup signal will be used as feedback for amplitude and phase correction, it is also important to know the phase of pickup signal w.r.t. to the field on axis. A variation in the RF phase from input RF port to the pickup loop is shown in Fig.17. The variation in the pickup phase with frequency is also studied considering a frequency drift of the cavity during operation due to thermal effects. The phase drift of pickup signal w.r.t. resonant frequency drift is same as that on the axis of the coupler cell as shown in Fig.18.

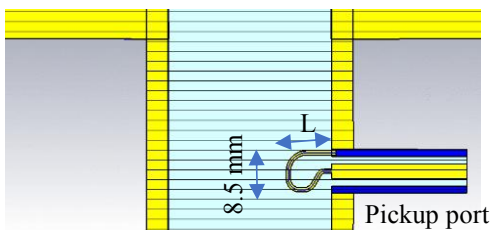

Fig.14: Model of pickup loop.

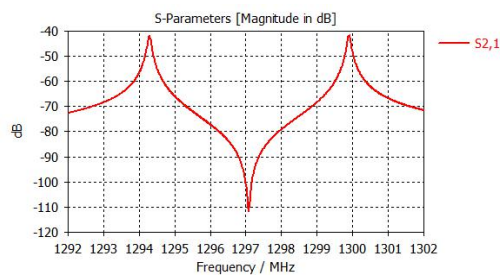

Fig.15: $S_{21}$ for pickup loop size of 8.5 mm x 15 mm.

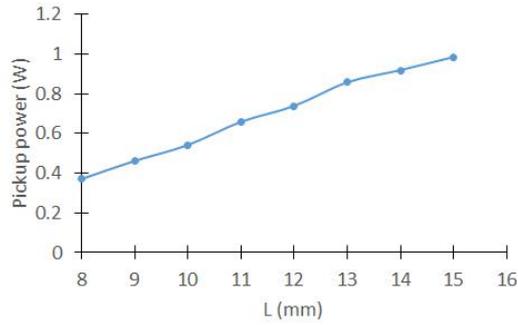

Fig.16: Variation in the pickup signal with the length of the pickup loop for loop width of 8.5 mm.

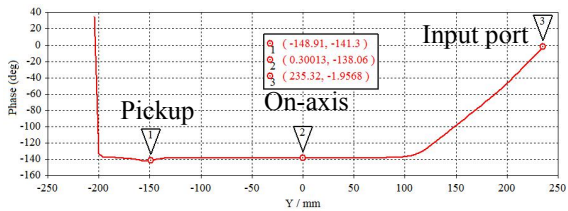
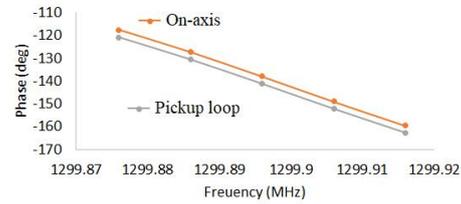

Fig.17: Variation in the phase from input port to the pickup loop.

Fig.18: Variation in the phase with frequency.

### 3.5. Coupler kick

Ideally, the transverse electromagnetic field on the beam axis in the buncher (or in any RF cavity in general) should be zero. However, the presence of the RF coupler port makes the transverse field on the beam axis non-zero in the coupler cell, which imparts a transverse kick to the beam. In-order to minimize the kick, the coupler cell is symmetrized by adding a vacuum and two dummy ports as discussed in previous section. In-spite of adding different ports to make the coupler cell symmetric, the transverse field on the beam axis is non-zero due to the phase asymmetry discussed in 3.3. The transverse fields on the beam axis calculated by CSTMWS is shown in Fig.19. The transverse momentum in the horizontal and vertical directions are calculated as [13]

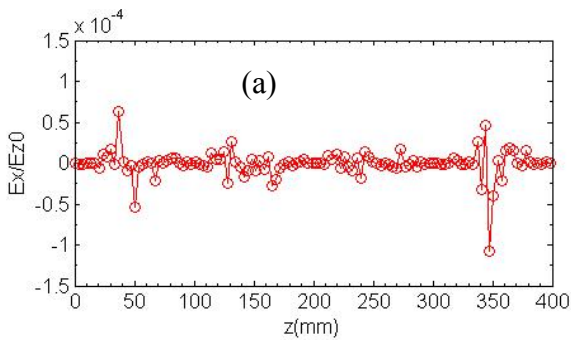
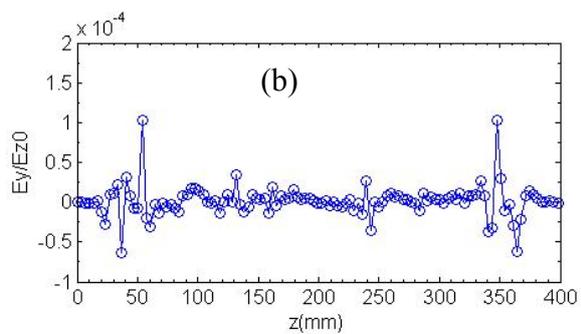

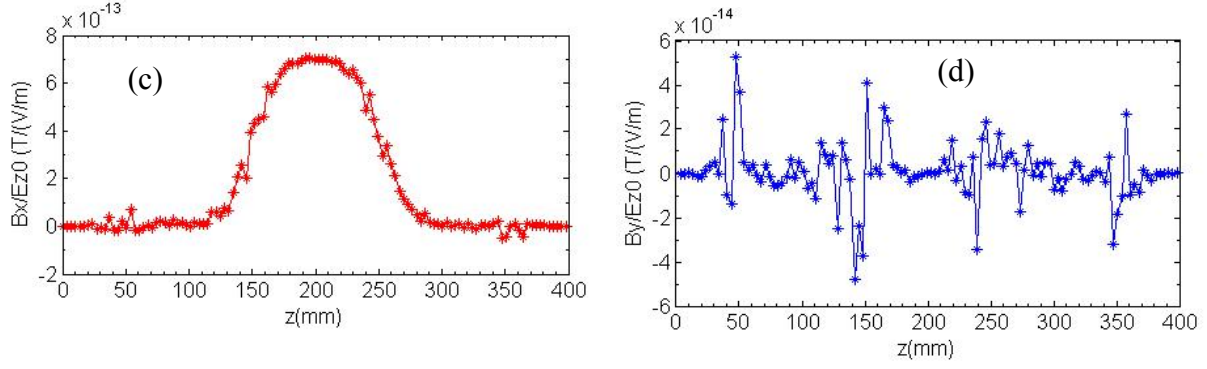

Fig.19: Normalized transverse fields: (a) Ex, (b) Ey, (c) Bx and (d) By on the beam axis of the buncher as simulated by Frequency domain solver of CSTMWS.

$$p_x = e \int_0^{L/\beta c} \left(E_x \sin(\omega t + \varphi_0) - v_z B_y \cos(\omega t + \varphi_0)\right) dt, \quad (14)$$

$$p_y = e \int_0^{L/\beta c} \left(E_y \sin(\omega t + \varphi_0) + v_z B_x \cos(\omega t + \varphi_0)\right) dt \quad (15)$$

where $E_{x,y}$ and $B_{x,y}$ are the electric and magnetic field on the beam axis. Variation of transverse momentum in horizontal (x) and vertical (y) directions with RF phase is shown in Fig.20. The maximum horizontal momentum is 1.19 eV/c while vertical momentum is 24.19 eV/c. The vertical kick is very close to the prediction as shown in ref [26]. Due to the symmetry of the cavity, there should be zero dipole kick along the horizontal plane, so it is probably numerical noise.

As shown by Eq. (13), the emittance growth induced by such a kick is ~0.001 mm.mrad without considering space charge effect, in principle negligible. Such a kick can be cancelled by using a pair of symmetric coupler ports, i.e. the vacuum port will be a power coupling port too. Since the predicted emittance growth is negligible, we want to avoid the complexity of a dual RF feed. In order to make sure such a kick does not increase emittance in cases with space charge, beam dynamics tracking with space charge and 3D buncher field maps will be presented in the following section.

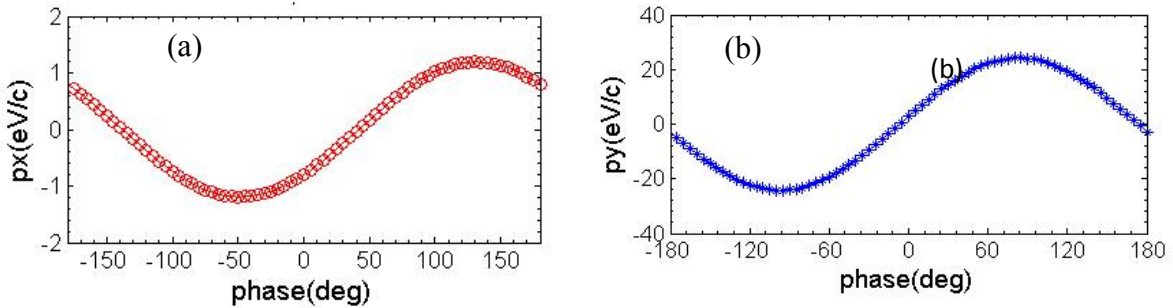

Fig.20: Variation of momentum with phase: (a) horizontal and (b) vertical.

## 4. Beam dynamics study

The effect of different ports on the buncher, on the beam parameters viz. the normalized transverse emittance and the RMS beam size at the end of the injector are determined by beam dynamics simulations using ASTRA [27] with 3D field of the buncher simulated from frequency solver of CSTMWS. The beam from the cathode is modelled as an ensemble of 10,000 micro-particles with a truncated Gaussian distribution in the transverse direction and a flattop distribution in the longitudinal direction. The bunch charge is 100 pC. The electromagnetic field in the buncher is scaled to have a cavity voltage of 400 kV. For different configurations of the buncher, the acceleration field amplitude and phase in the buncher are matched, and all other beamline settings are the same. The evolution of the normalized transverse emittance and the rms beam sizes in the horizontal and vertical direction versus the distance from the cathode is shown in Fig.21 and Fig.22 respectively. A summary of the normalized transverse emittance for different configurations of the buncher is given in the Table 8. From Table 8 it is clear that presence of the four ports, the emittance and the beams sizes are very close to No-port case. Without the dummy port to compensate the quadrupole component in the buncher coupler cell, a ~16% emittance growth is observed. Besides, the phase slope shown in Fig. 17 due to the single RF port did not perturb the beam emittance and symmetry as expected by analytical analysis, and a single RF port reduced the waveguide complexity around the buncher cavity.

Table 8: Comparison of emittance for different configurations of the buncher.

| Configuration | No-ports | Two ports (RF and vacuum) | Four ports (RF, Vacuum, dummy) |
|---|---|---|---|
| $\varepsilon_x$(mm-mrad) | 0.19162 | 0.22455 | 0.19979 |
| $\varepsilon_y$(mm-mrad) | 0.19621 | 0.22695 | 0.20387 |

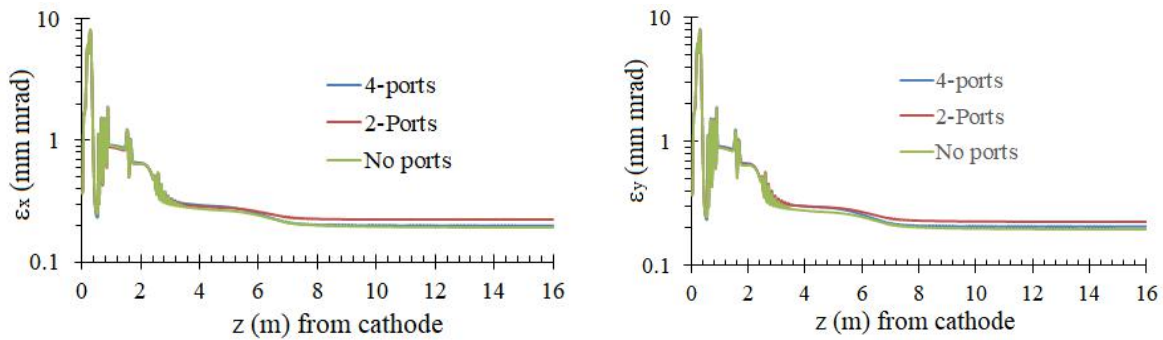

Fig. 21: Evolution of normalized rms emittance in the horizontal and the vertical direction for different configurations of the buncher.

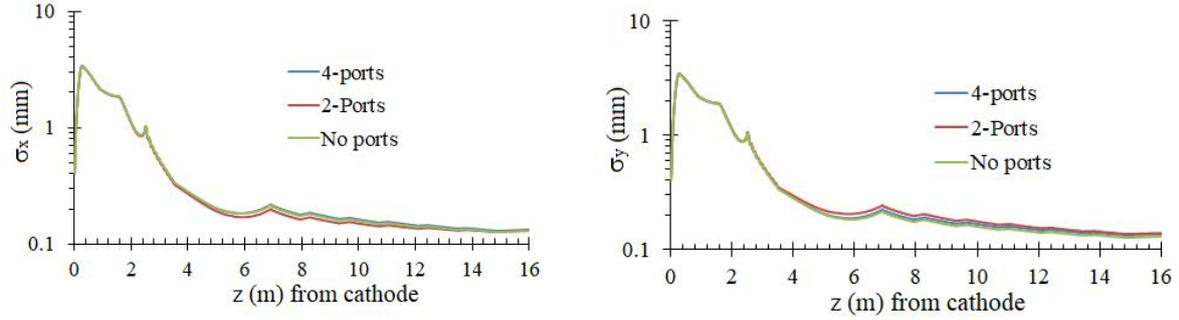

Fig. 22: Evolution of rms beam size in the horizontal and the vertical direction for different configurations of the buncher.

## 5. Mutipacting Study

For smooth operation of an RF cavity, it must be mutlipacting (MP) free around the operating power level. The multipactor effect is a phenomenon in RF structures, where, under certain conditions, secondary electron emission is in resonance with the RF field leading to exponential electron multiplication, possibly perturbing the RF and even damaging the RF structure [28]. The strength of MP is determined by growth rate of the electrons with time by fitting as $N(t) = N_0 e^{\alpha t}$, where $N_0$ is the number of primary electrons, and α is the growth rate expressed as ns$^{-1}$. A positive value of α indicates the presence of MP, however, it is observed that α< 0.05ns$^{-1}$ is also safe for operation of a cavity [29]. Therefore, the cavity has to be optimized to have α < 0.05 ns$^{-1}$ for operating voltages.

The multipacting (MP) study of the buncher is carried out using Particle-In-Cell (PIC) solver of the CST Particle Studio (CSTPS) [18]. For multipacting simulations CST utilizes a hexahedral mesh, which has poor accuracy as compared to a tetrahedral mesh and is unable to accurately represent the small features of the structure such as the nose cone or inter-cell coupling iris etc. where the probability of multipacting is high. Although the accuracy of simulations can be improved by increasing the mesh numbers, however it also increases the simulation time and limited by the computer memory. To have a good accuracy within a reasonable simulation time, the RF fields from the Eigen mode solver with tetrahedral meshing are saved as an ASCII file and then imported for multipacting simulations with proper scaling [29]. Since multipacting occurs near the surfaces, the mesh density in the Eigen mode solver is enhanced near surfaces. The cavity is meshed as two shells of vacuum as discussed in detail in reference [30]. The outer-shell is modeled with a thickness of the order of 1-2 mm with higher mesh density. A typical meshing view of the two-shell model of the cavity is shown in Fig.23. With this approach the accuracy of the electromagnetic field near the surface is improved. After that the meshing is refined for PIC model in steps to reach convergence of the MP growth rate at a particular operating voltage. After fixing the mesh the cavity voltage is scanned to find the MPs.

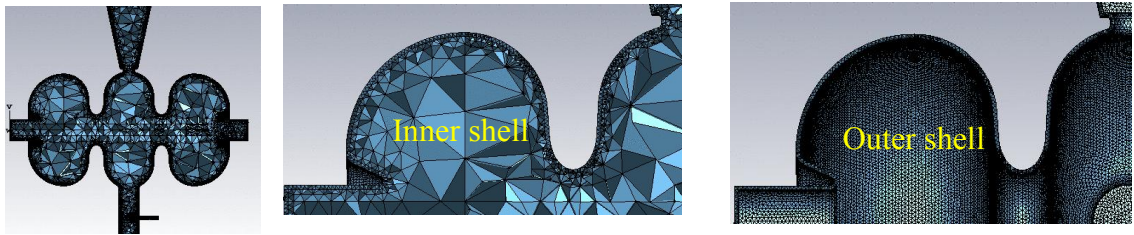

Fig.23: Meshing view of the three cell buncher used in Eigen mode solver to calculate electromagnetic field: showing two shell model with finer mesh near surface.

The MP simulation also critically depends upon the Secondary Emission Yield (SEY) of the material (copper in our case), which depends upon the composition and processing of the material. In literature SEY (maximum) of copper is reported in the range of 1.2 to 2.3 with different treatments as shown in Fig.24 [31]. Although the composition and processing of copper to be used for fabrication of the buncher is unknown at this time, however the structure is sure to be brazed. The typical brazing temperature is ~ 800° C or more depending upon the brazing material. The SEY of copper for the case 'oxidized and backed at 350° C' with a maximum SEY of 1.4, is used in Ref [30] where experimental results agree very well with the simulations. Considering the uncertainty in SEY and to be on safer side, we checked the MP of the buncher for SEY curves with peak values of both the 1.4 (baked and oxidized at 350°C) and the 1.87 (backed at 300 deg C). The variation of α with accelerating voltage for different SEY data is shown in Fig.25. For both the SEY curves, there is no multipacting up to the cavity voltage of 400 kV.

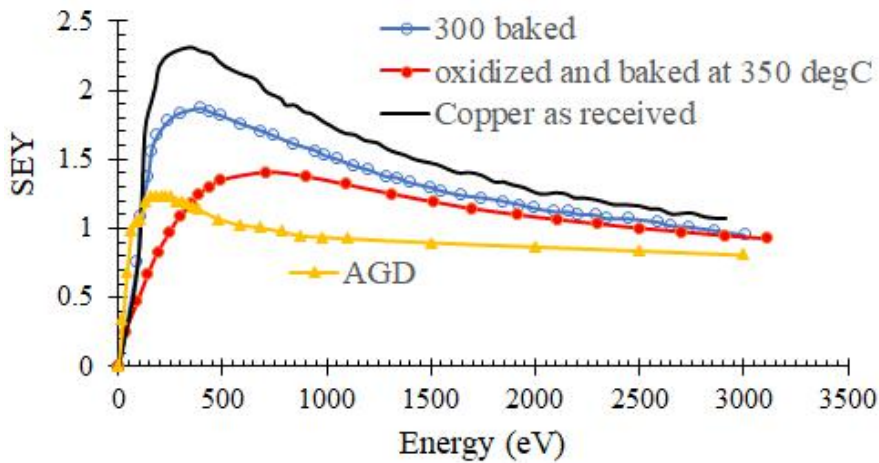

Fig.24: Comparison of SEY of copper after different processing (data taken from Ref [31]).

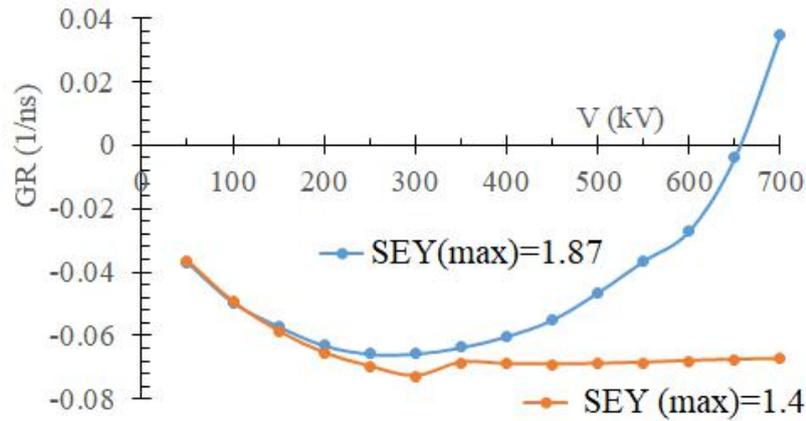

Fig. 25: Particle growth rate with accelerating voltage in the three cell buncher predicted by PIC solver of CSTPS.

## 6. MULTIPHYSICS ANALYSIS

Dissipation of ~14 kW CW RF power causes cavity temperature rise, structure deformation, thermal stresses and cavity frequency shift. In order to ensure a reliable operation of the buncher at rated RF power and to remove the RF heating a water-cooling scheme is designed. A coupled RF-thermal-mechanical analysis of the buncher is carried out using Multiphysics solver of CSTMWS [18]. For coupled simulations, the thermal load is evaluated using electromagnetic field distribution from Eigen mode solver and exported to the thermal solver. Then, the thermal load is scaled to the required cavity voltage of 400 kV and the steady-state temperature distribution is evaluated using thermal solver and exported to the structure solver, where stress and structural deformations are calculated. After that the structural deformations are exported to the Eigen mode solver to evaluate the frequency detuning. A flow chart of the coupled RF-thermal-mechanical Multiphysics simulations is shown in Fig. 26.

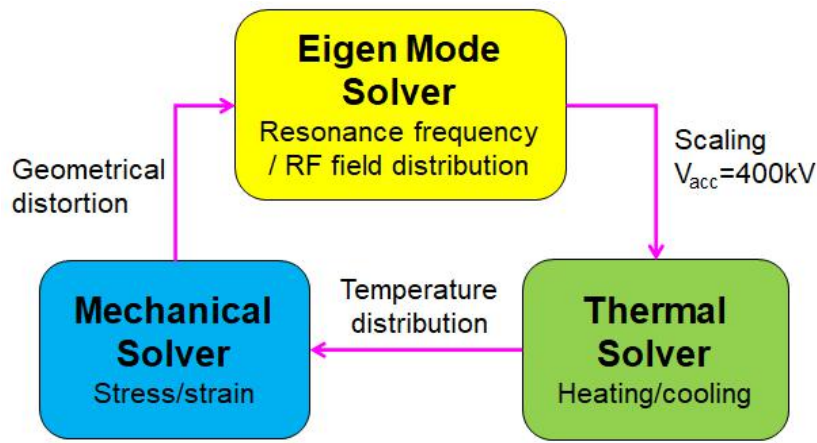

Fig. 26: Flow chart for Multiphysics simulations.

## 6.1. Thermal analysis

Based on electromagnetic field distribution, water cooling channels are modelled near the nose cones, ID of the end cells, inter-cell coupling iris and the ID of coupling cell. A solid model of the three cell buncher with cooling channels is shown in Fig. 27. The body of the buncher is tapered at end wall, considering the resonance frequency tuners of the end cells. Considering the machining feasibility and possible brazing locations, the cooling channels near the end walls and near the inter-cell coupling iris are made of circular cross-section with a diameter of 10 mm and square path like used in BNLphotocathode RF gun [33] while cooling channels near the cell IDs are made with a rectangular cross-section of 5 mm x10 mm with circular path. Since the thermal solver of CST takes the heat transfer coefficient, temperature of cooling water and ambient temperature as input parameters to perform the thermal analysis, hence these parameters are calculated analytically. As discussed in Ref [34], the heat transfer coefficient and pressure drop are derived using the following empirical formulas

$$h = \frac{\lambda}{D} N_u, \tag{16}$$

$$N_u = 0.023 Re^{0.8} Pr^{0.4}, \tag{17}$$

$$Re = \frac{\upsilon D \rho}{\mu}, \tag{18}$$

where $\lambda$ is thermal conductivity of the fluid (water 0.58 W/mK), $\upsilon$ is liquid velocity (m/s), $\rho$ is liquid density (water 1000 kg/m³), $\mu$ is liquid viscosity (water ~0.8e-3 kg/ms), Re is the Reynold number, Nu is the Nusselt number, Pr is the Prandtl number (5.7 for copper), D is hydraulic dimeter of cooling channel (Area/perimeter) and h is heat transfer coefficient (W/m²K). The Pressure drop is calculated using the Darcy Weisbach equation

$$\Delta p = f_D \frac{L}{D} \frac{\rho v^2}{2}, \tag{19}$$

where $\Delta p$ is pressure drop (N/m²), $f_D$ is the Darcy friction factor, and L is pipe length (m) [34]. The heat transfer coefficient and pressure drop for the circular and the rectangular shaped cooling channels are summarized in Table 9.

| Table 9:Pparameters of the water-cooling channelsused in thethermal simulation. | | | | | | |
|---|---|---|---|---|---|---|
| Cooling pipe cross section | Dimensions | Velocity (m/s) | Flow rate (liter/s) | Re | H (W/m²K) | Δp (kPa) |
| Circular | 10 mm (diameter) | 3 | 0.23 | 29441 | 10062 | 0.31 |
| Rectangular | 10 mm x 5 mm | 3 | 0.15 | 19627 | 10912 | 0.17 |

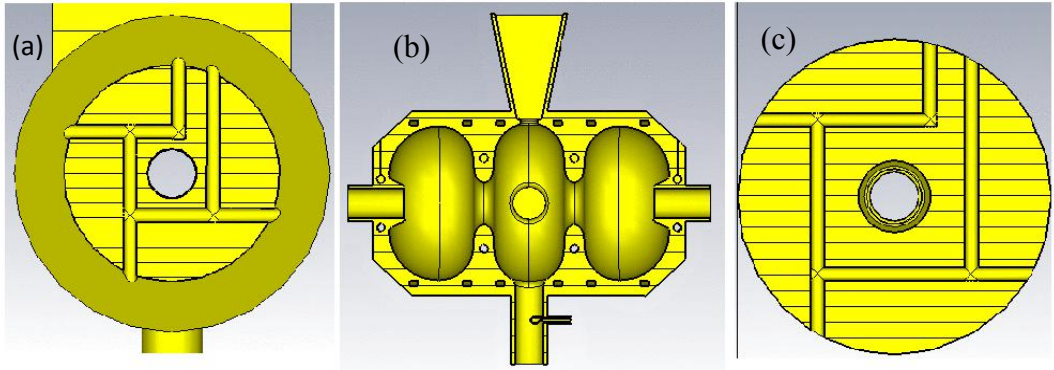

Fig.27: Schematic view of the buncher showing different cooling channels: (a) near the end wall and (b) cross-section of cooling channels (c) at inter-cell coupling iris.

The convective boundary conditions are applied to the surfaces of the cooling channels and air with normal flow is used as background material with a space of 2 mm from the outer surface (copper) of the buncher. Since the typical ambient temperature of the Eu-XFEL tunnel is 25°C, the cooling water temperature is also taken as 25°C. The steady state temperature distribution in the buncher is shown in Fig. 28. The maximum temperature is 48.9°C, which is near the RF coupling slots as shown in Fig. 28, and is well below the acceptable limits, indicating that the cooling is sufficient.

### 6.2. Structural analysis

For structure simulation, the temperature distribution from thermal solver is imported into the structural solver. The von Mises stress distribution in the buncher structure is shown in Fig.29 while geometrical deformation map is shown in Fig. 30. The maximum displacement is <15 μm at the ID of end cells while it is <35 μm at the ID of coupler cell. The peak stress is of the order of 26 MPa which is near the RF coupling slot. The maximum stress is much lower than the yield strength of the soft copper of 62 MPa [36]. This is again showing that the cooling is sufficient.

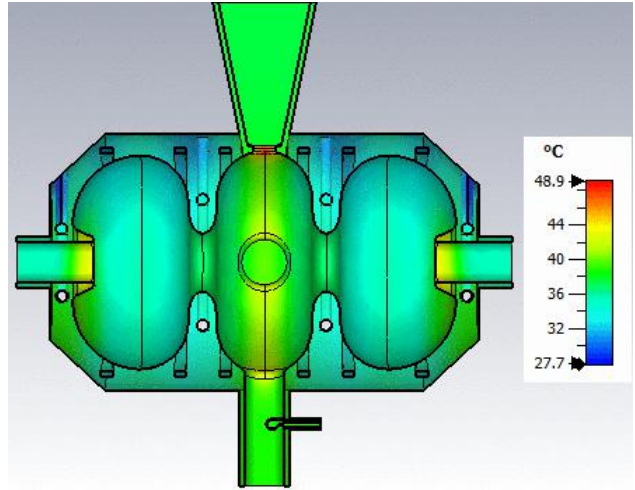

Fig.28: Temperature distribution in the buncher for RF power dissipation of 14 kW and cooling water of 25°C supplied at 3 m/s.

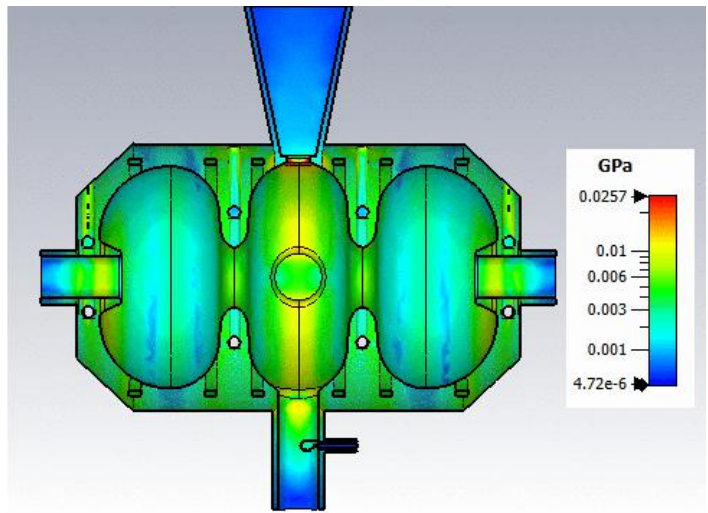

Fig.29: Stress distribution in the three-cell buncher for RF power dissipation of 14 kW (accelerating voltage 400kV) with cooling water of 25°C supplied a speed of 3 m/s.

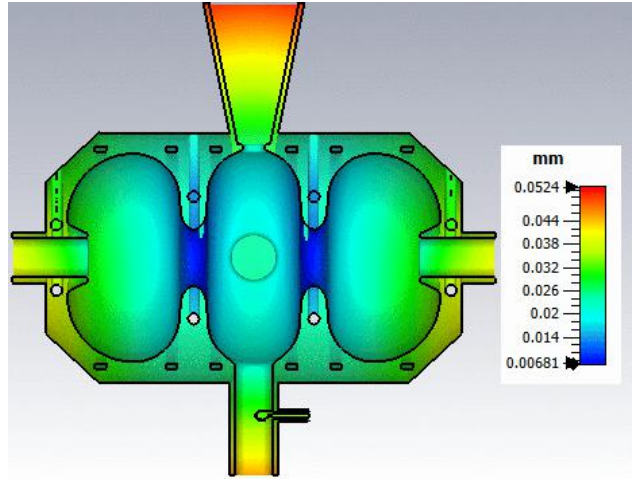

Fig.30: Displacement distribution in the three-cell buncher due to thermal expansion.

### 6.3. Frequency de-tuning due to thermal load

To determine the frequency detuning due to thermal deformations, the structural displacement from the structural analysis is exported to the Eigen mode solver of the CST MWS under the sensitivity analysis. The frequency shift of the π mode is -250 kHz due to structural deformations caused by the RF power dissipation. To compensate the change in the π mode frequency, the IDs of cells are modified to have π mode frequency of 1300.250 MHz (in vacuum with ambient temperature of 25°C) so that during operation (feeding of RF power of 14 kW and cooling water of 25°C flowing with a speed of 3 m/s) it becomes the target value of 1300 MHz. A summary of the ID's of the cells and the resonance frequency is given in Table 10. The change in cell ID's are within the manufacturing tolerances. The fine tuning of the frequency can be done by varying the cooling water temperature. The typical variation in the π mode frequency with cooling water temperature is shown in Fig. 31.

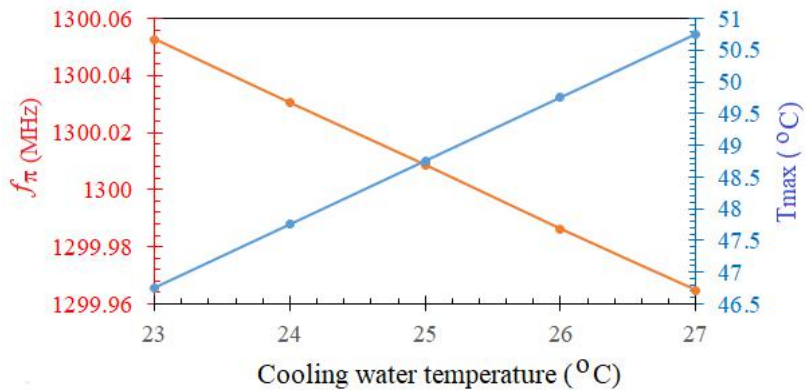

Fig.31: Variation in the π mode frequency and maximum temperature in the three cell buncher for RF power dissipation of 14 kW and cooling water flowing at 3 m/s.

Table 10: Summary of cell ID and π mode frequency before and after tuning.

| Parameters | Original | Revised |
|---|---|---|
| ID of coupler cell (mm) | 99.350 | 99.330 |
| ID of end cells (mm) | 95.241 | 95.220 |
| $f_\pi$ (MHz) | 1300.00 | 1300.25 |
| $f_\pi$ (MHz) with thermal distortions and cooling water of 25°C following at 3 m/s | 1299.750 | 1300.00 |

### 6.4. Transient thermal analysis

As discussed above, the frequency drift due to thermal load at steady state can be pre-compensated by adjusting the ID's of the cells during fabrication. Before the buncher cavity reaches the thermal equilibrium, the resonance frequency ($f_\pi$) of the buncher needs to be matched to the drive frequency to minimize the RF reflection. This can be done either by varying the drive frequency to match the buncher resonance frequency or by tuning the buncher frequency by varying the cooling water temperature. Since any variations in the cooling water temperature take some time to stabilize the resonance frequency of the buncher while driving frequency of the RF source can be done very fast, which is usually used in startup of high average power cavities now days. The variation in the buncher frequency with time is simulated by transient Multiphysics analysis of the buncher using Multiphysics solver of CSTMWS. First the variation ofcavity temperature with time is determined by transient thermal analysis, where the nominal thermal load of 14 kW from Eigen mode solver is applied with a rectangular signal of 500s. The variations in temperature at three different locations (1) RF coupling slot, (2) vacuum port and (3) ID of the coupler cell are shown in Fig.32. The temperature reaches steady state in ~200s. For structural analysis, the temperature distribution in the thermal solver is saved at different time intervals and exported to the structural solver to determine the structural deformation with time. The structural deformation at different times is then exported to the Eigen mode solver to determine the frequency shift with time. A variation in π mode frequency with time for the input power of 14 kW (cavity voltage =400 kV), is shown in Fig.33, which shows the cavity frequency stabilized after 100 seconds.

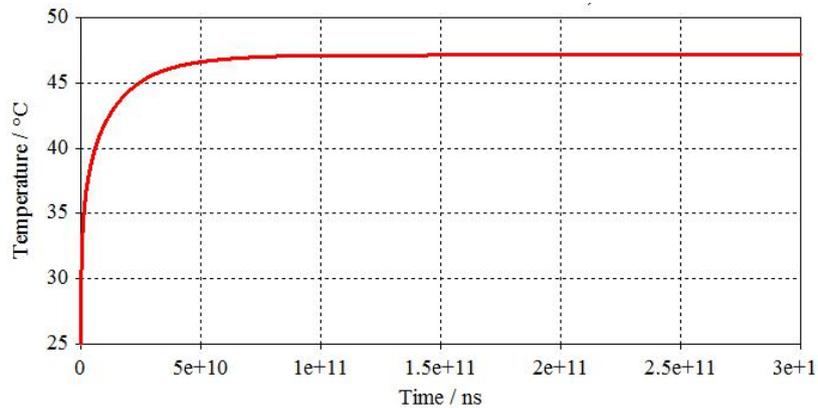

Fig.32: Variation in temperature with time at different locations in the buncher.

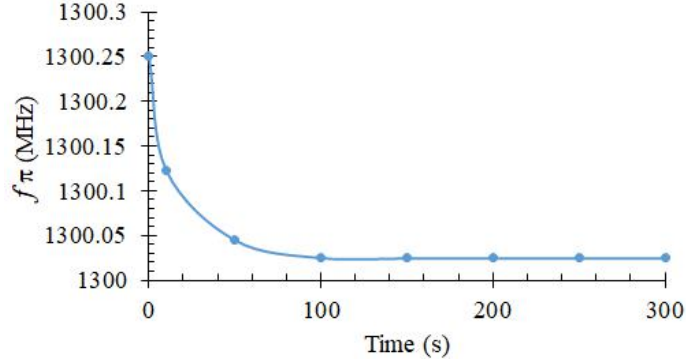

Fig.33: Variation in resonance frequency with time.

## 7. Tolerance and tuning methodology

In general, even a structure is designed with very fine mashing, the machining tolerances during fabrication and brazing cause a deviation from the design values, and the cavity needs to be tuned. In our case the main concern is the $\pi$ mode frequency and field flatness. Generally, the tuning is done through an iterative cut-and-measure or deform-and-measure procedure involving small modifications in dimensions of the individual cells followed by RF measurements, till the desired $f_\pi$ and field flatness are achieved. Since machining cut is not possible after brazing of the structure hence deform-and-measure procedure is considered for tuning. In order to determine the required tuning range, a tolerance study is carried out where geometrical dimensions of the buncher as shown in Fig.34, are varied one by one and variation in resonance frequency, field flatness, quality and shunt impedance are calculated using Eigen mode solver of the CSTMWS. Since with a reasonable machining facility, dimensions can be achieved with ±10μm, the geometrical dimensions are varied within this range and variation in the resonance frequency is fitted as following equations. It is noted that the variation the quality factor, field flatness, RF coupling coefficient are negligible.

$$\Delta f_\pi(kH_z) = -14.23\ \Delta R_{c2}(\mu m) + 6.14\Delta L_2(\mu m) - 5.2\Delta a(\mu m) + \Delta b(\mu m) - 7\Delta A(\mu m) + 2.5\Delta B(\mu m) \quad (20)$$

$$\Delta f_\pi(kH_z) = -14.20\ \Delta R_{c1}(\mu m) + 6.5\Delta L_1(\mu m) - 4.2\Delta R_5(\mu m) - 2.5\Delta R_3(\mu m) \quad (21)$$

If we take a worst case where all dimensions deviate such that the change in frequency is in same direction, then the maximum deviation in $f_\pi$ will be ~360 kHz. The brazing effect is usually hard to predict, therefore it is not discussed here.

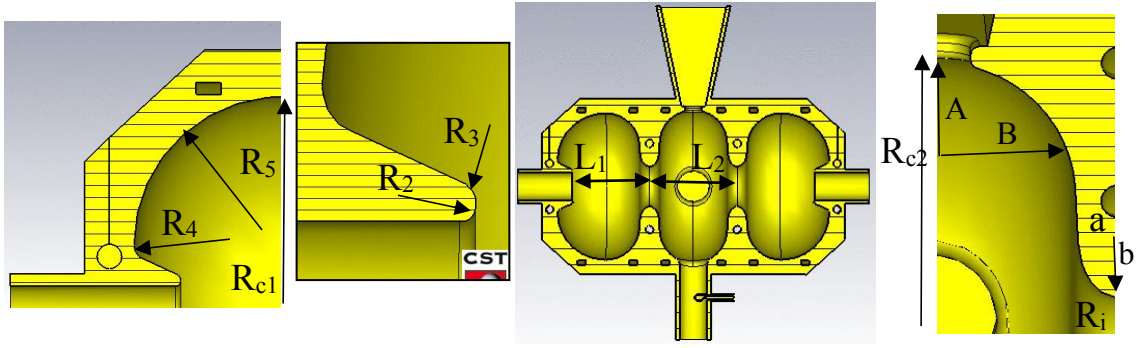

Fig.34: Parametric view of the three cell buncher cavity.

### 7.1. Tuning methodology

The frequency $f_\pi$ can be tuned by deforming the cavity wall locally by pushing or pulling [37]. To eliminate the possibility of excitation of dipole or quadrupole mode, we planned to have four tuners diametrically opposite and orthogonal to each other in each cell as shown in Fig.35. Since CSTMWS do not take plastic deformation of material in the structural solver therefore change in frequency is evaluated by using Slater's perturbation theorem analytically [38]. If the change in volume is '$dv$' due to wall deformation then the change in resonance frequency is given by [38]

$$\frac{\Delta f_\pi}{f_\pi} = \frac{\int (\mu H^2 - \varepsilon E^2) dv}{\int (\mu H^2 + \varepsilon E^2) dV} = \frac{\int (\mu H^2) dv}{4U}, \qquad (22)$$

where U is the energy stored in the cavity, H and E are the magnetic and electric field at the location of tuner. Since in present case, at the location of tuner, electric field is 'zero' hence only the magnetic field term is applicable. If an ellipsoidal approximated deformation volume at the tuner location is $dv = (2/3)\pi abc$, where 'a', 'b' and 'c' are the semi-axis of the ellipsoidal. For an ellipsoidal of having two semi-axes of 12.5 mm, the variation in the frequency with third semi-axis (depth of deformation) is given in Fig.36. With a deformation of $<\pm 2$ mm, frequency can be tuned to $\pm 100$ kHz/tuner which is more than the expected frequency deviation due to machining errors.

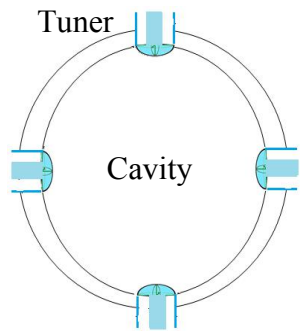

Fig.35: Schematic view of tuners.

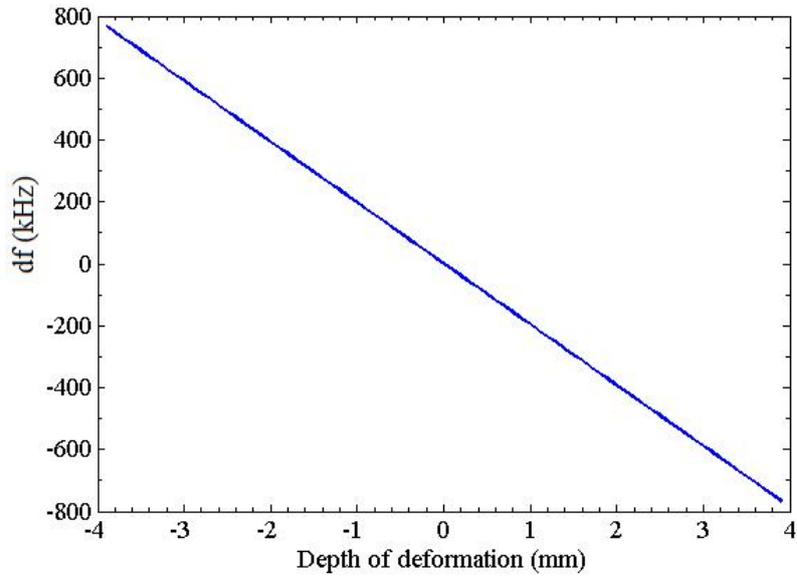

Fig.36: Variation in the resonance frequency with depth of deformation.

## 8. Conclusion

RF design of a normal conducing 1.3 GHz CW buncher for a VHF gun-based injector for the European X-FEL is carried out using CST MWS. The geometries of the buncher are optimized to achieve a higher mode separation as well as higher shunt impedance. A thermal load of ~4.7 kW/cell is achieved for a nominal cavity voltage of 400 kV, which is roughly a factor of 2 improvement over existing buncher designs. A tapered waveguide-based RF power is designed to feed RF power in to the buncher. The coupler cell is optimized to minimize the field asymmetry induced coupler kick. The emittance growth due to coupler cell asymmetry is minimized and verified with the beam dynamics simulations. The multipacting simulations were carried out using CST-PIC and the buncher is MP free up to cavity voltage of 600 kV. The mechanical deformation due to RF heating is carried-out using Multiphysics solver of CST MWS and a cooling circuit is designed to minimize the thermal load effect. A cavity deformation methodology is also evaluated to tune the buncher frequency.

Annexure -1

**Velocity acceptance of two and three-cell bunchers**

Although the buncher is designed assuming the input beam energy of 850 keV from a VHF gun, however the optimization of the VHF gun is still underway. Since the input beam to the buncher is non-relativistic, a variation in its energy causes a variation in the electron's velocity. Variation in the electron's velocity leads to vary the transient time factor (TTF), defined as

$$TTF = \frac{\left|\int_0^L E_z(z) e^{-i\frac{2\pi z}{\beta\lambda}} dz\right|}{\int_0^L |E_z(z)| dz}. \qquad (23)$$

Variation in TTF varies the effective accelerating voltage (for a fixed input RF power), and variation in the actual energy gain due to velocity (energy) mismatch over maximum energy gain is known as velocity (energy) acceptance of the structure. Since in a π mode, multi-cell RF accelerating structure, the effective cavity voltage is maximum when the beam is synchronized with the RF in each cell (time taken by the beam to travel from one-cell to next cell and phase advance of the RF in that time is π). Generally, the shunt impedance of an RF accelerating structure increases with the number of cells, however its velocity acceptance starts reducing. Variation in TTF for the two and the three-cell is shown in the Fig.37. For both the cases TTF is maximum for β ~1. The TTF is determined by calculating the effective shunt impedance using postprocessor of the Eigen mode solver of the CSTMWS. The relative reduction in the TTF for the two-cell buncher is 2.8 % when the input beam energy reduces from 1MeV to 750 keV, while it is 2.3 % for the three-cell buncher. Although there is a slight reduction of the velocity acceptance for the three-cell bucher case, the total shunt impedance for the three cell buncher is ~40 % higher as compare to the two-cell buncher (see Table 1 and Table 2), therefore using the three-cell bucher is more suitable.

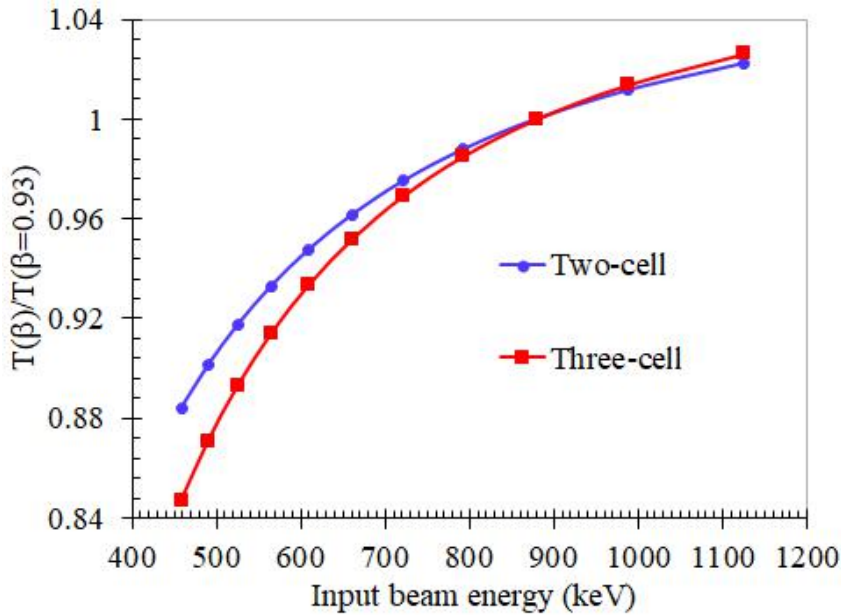

Fig.37: Variation of the normalized shunt impedance with the input beam energy of the two and three-cell bunchers.

Annexure -2

**Alternativecoolingscheme**

An alternative cooling scheme is also studied. In this scheme the body of the buncher is taken a solid square as shown in Fig.38. The body is tapered at end walls and four diagonally orthogonal locations to have a feasibility to machine tuners for tuning the frequency of buncher cells. All the cooling channels are modeled with a circular cross-section with diameter of 10 mm. The channels near the end walls and near the inter-cell coupling iris are made square path like LCLS photocathode RF gun [32] while cooling channels near the cell IDs are drilled as shown in Fig.39. The advantage of this scheme is that the in this scheme there is no requirement of brazing the cooling channels.

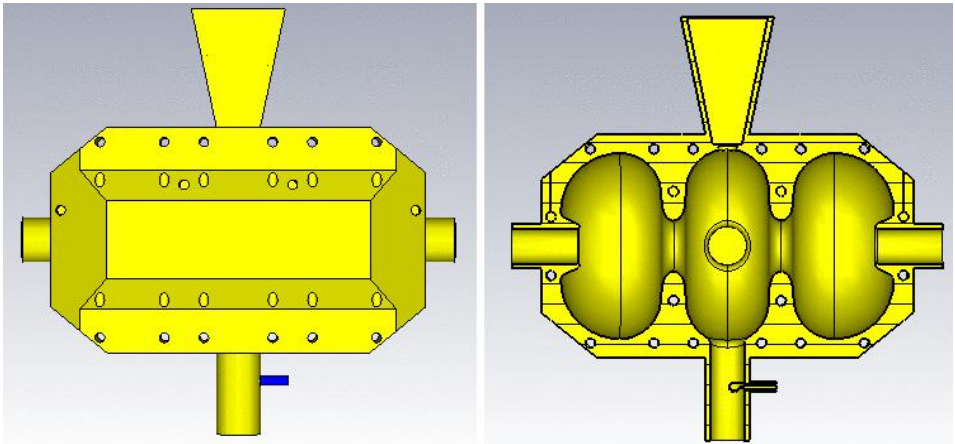

Fig.38: Schematic view of the buncher body with a cut section.

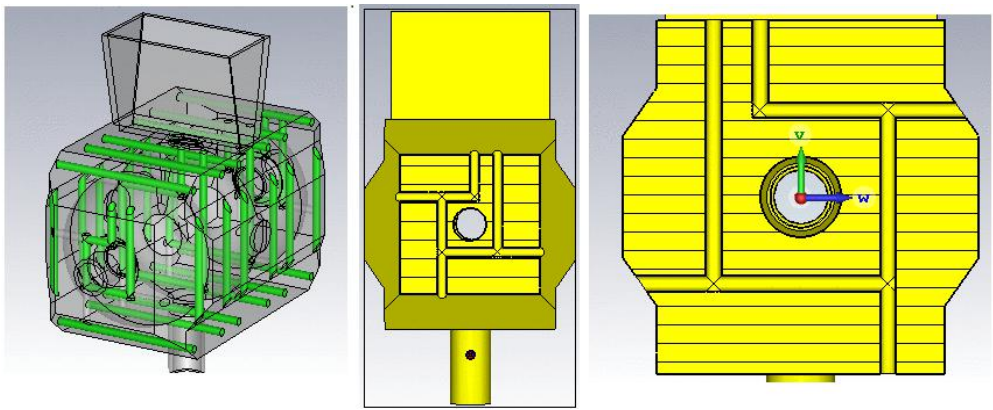

Fig.39: Schematic view of different cooling channels modeled in the buncher: (a) near the end wall and (b) at inter-cell coupling iris.

With the ambient temperature of 25°C and cooling water of 25° C flowing at 3 m/s, the steady state temperature distribution in the buncher is shown in Fig. 3 while stress distribution in shown in Fig.4. The maximum temperature is 47.6°C and the maximum stress is 28.8 MPa is near the RF coupling slots as shown in Fig. 40 and Fig.41 respectively. The maximum temperature and stress are within acceptable limits, indicating that the cooling is sufficient. The thermal stress causes a frequency shift of -0.247 kHz. The performance of the is scheme is similar to the cooling scheme in the main report.

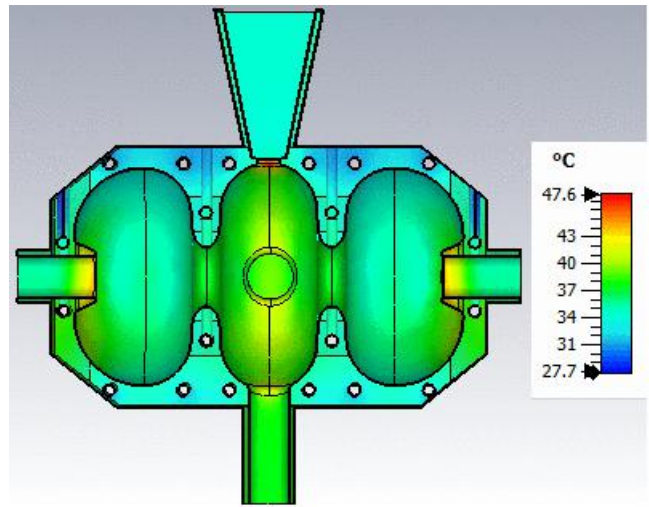

Fig.40: Temperature distribution in the buncher for RF power dissipation of 14 kW and cooling water of 25°C supplied at 3 m/s.

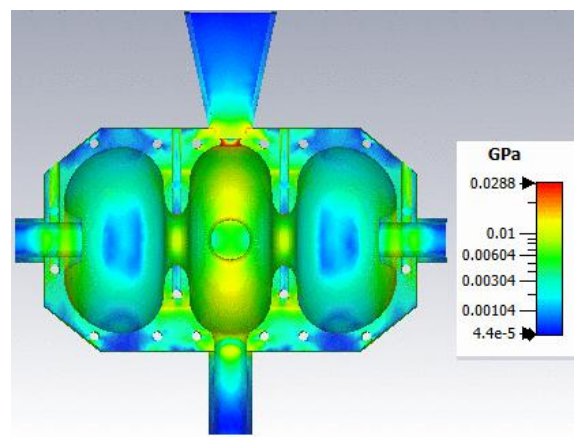

Fig.41: Stress distribution in the three-cell buncher for RF power dissipation of 14 kW (accelerating voltage 400kV) with cooling water of 25°C supplied a speed of 3 m/s.

# Appendix -A

Very through design study was performed and described in above Sessions. Including all steps of cavity design, the study above gave both a rationale for general decisions and revealed the mutual connections of particular decisions. Another cavity design, presented below, makes substantial use of the results of the previous study and at the same time is based on the mastered technological level and experience of manufacturing DESY L-band RF Gun cavities. The general view of the cavity and its constituent parts are shown in Fig. A1.

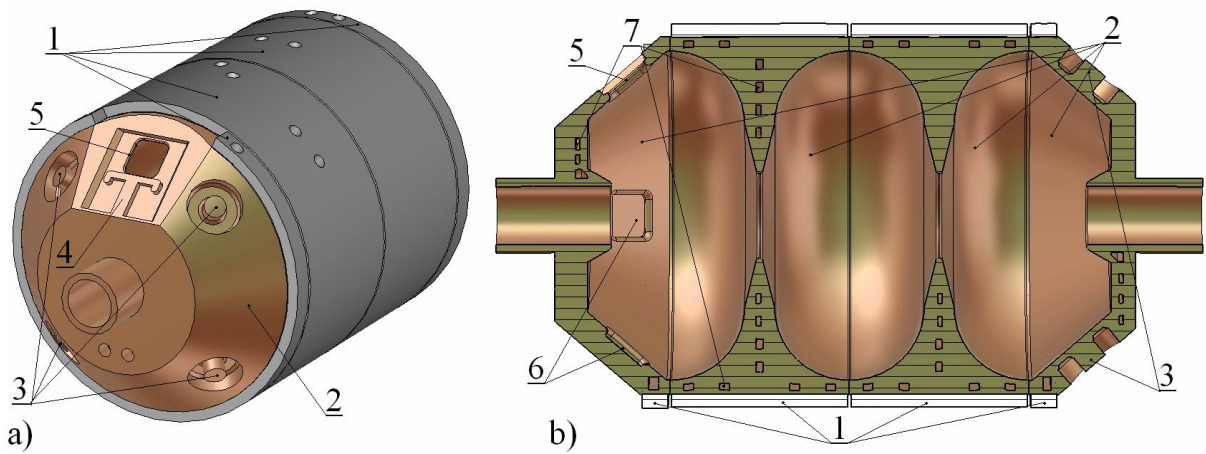

Figure A1. General view of the cavity and its constituent parts. 1- parts of stainless steel jacket, 2- copper cavity parts, 3- RF tuners, 4 – place for connection with tapered driving waveguide, 5 – matching RF window, 6 – blind holes for multipoles compensation, 7 – cooling channels.

## A1. RF cavity design.

Rationale for three cells cavity is shown in Sessions 2.2 and 2.3. Simultaneously, consideration in Table 1 shows the big reserve in the maximal electric field value $E_{smax}$ at the RF surface in considered CW bunchers.

## A1.1. RF cells design.

Conditions of low $E_{smax}$ value and RF efficiency are contradictory. Drift tubes noses and iris tips are optimized to improve slightly $Z_e$ value and, mainly, improve mainly, modes separation in frequency. The definitions of cells dimensions are illustrated in Fig. A2.

The considered cavity can be realized in two options – uniform RF losses and equal $E_z$ amplitudes. Let us to consider mainly the first option of uniform RF losses. The cavity cells have the same radius $R_c$, and are tuned for the same operating frequency f=1300 MHz with magnetic boundary conditions in the middle of irises. The middle cell is tuned by $R_c$ fitting. This $R_c$ value is applied to end cells and RF tuning is performed by $r_2$ fitting. The conical parts in the first and the third cells are foreseen for more comfortable placement of RF coupler, RF tuners and RF probes. The length value for end cells $L_2$ is selected from the maximal $Z_e$ value. The cone angle for drift tubes is selected as 30 degrees from conditions of RF efficiency and drift tube cooling. The cells dimensions in mm for option of uniform losses are listed in the Table A1 and calculated by using MWS CST [18] are listed in the Table A2. Calculated values of mode frequencies are $f_0$=1286.87 MHz, $f_{\pi/2}$=1295.384 MHz, $f_\pi$=1300.00 MHz.

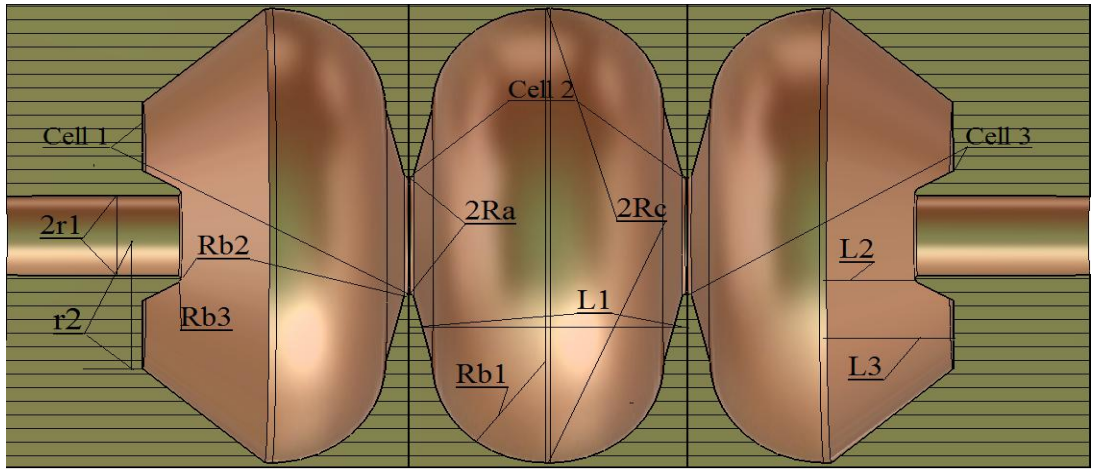

Figure A2. Definition of cells and cell sizes.

Table A1. Dimensions of the cavity cells for option of uniform RF losses.

| Parameter | r1 | r2 | Ra | Rc | Rb1 | Rb2 | Rb3 | L1 | L2 | L3 |
|---|---|---|---|---|---|---|---|---|---|---|
| Value, mm | 18.0 | 41.532 | 26.0 | 100.966 | 44.75 | 1.0 | 2.5 | 109.5 | 35.0 | 15.0 |

Table A2. Calculated RF parameters of the cavity for option of uniform RF losses

| Parameter | Value |
|---|---|
| Frequency, $f_\pi$, MHz | 1300.00 |
| Quality factor, $Q_0$ | 27710 |
| $Z_e$, M$\Omega$ | 12.68 |
| Mode separation $f_\pi$-$f_{\pi/2}$, MHz | 4.6 |
| Power dissipation Pc (kW) for 400 kV | 12.62 |
| $E_{smax}$, MV/m for 400 kV | 10.13 |
| $E_{smax}/E_k$ | 0.32 |

As it is mentioned, the cells dimensions, specified in the Table A1, corresponds to the option of uniform RF losses, which is illustrated in Fig. A3a. As one can see from Fig. A3a, distribution of RF loss density along cavity surface, except a natural enhancement near drift tubes, is rather uniform.

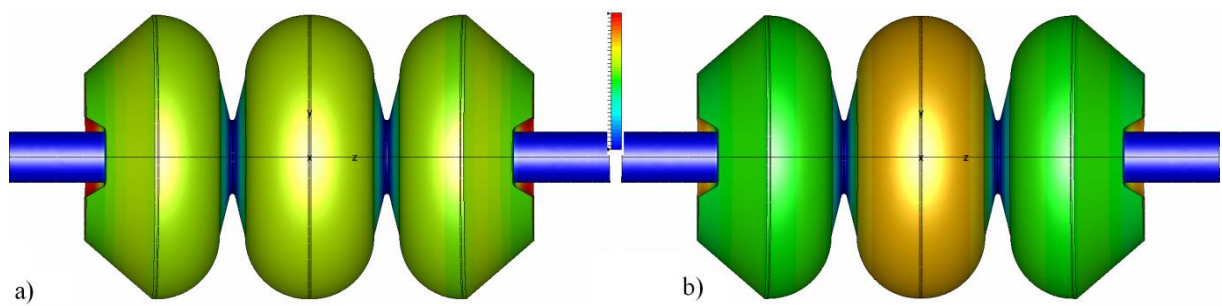

Figure A3. RF loss density distribution for cavity option with uniform RF loss distribution, (a) and option with equal Ez amplitudes.

But this option has reduced amplitude of Ez component, as it is shown in Fig. A4a. Distribution of Ez component depends on cells frequencies. With reducing the outer radius of the middle cell to 100.869 mm, so increasing the frequency of the middle cell, and appropriate increasing the outer radius of end cells to 101.004 mm, to keep operating frequency f=1300.00 MHz, we easy get equal amplitudes Ez component in cells, see Fig. 4b. It corresponds to RF loss density increasing in the middle cells, see Fig. A3b and a small reduction in Ze value to 12.5 Mom.

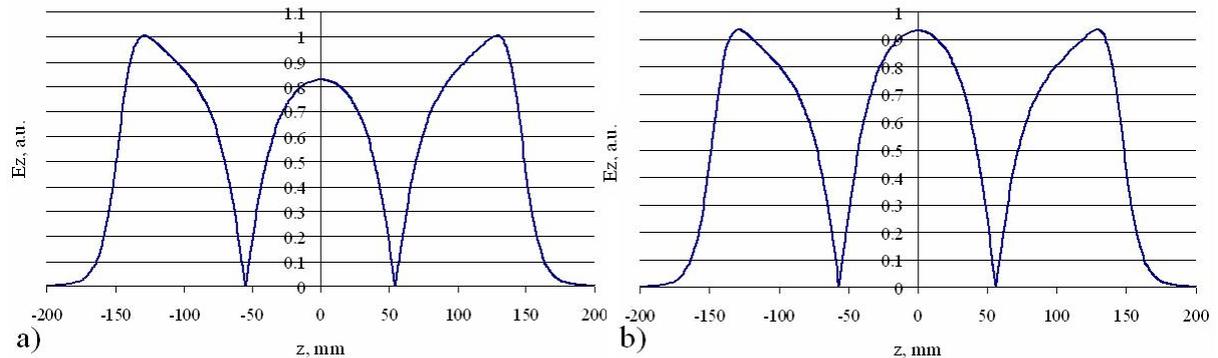

Figure A4. Plots of |Ez| component along cavity axis for cavity option with uniform RF loss distribution, (a) and option with equal Ez amplitudes.

From comparison of cells dimension for two cavity options we see conclusion – even with improved modes separation field distribution along cavity axis is sensitive enough to the balance of cells frequencies.

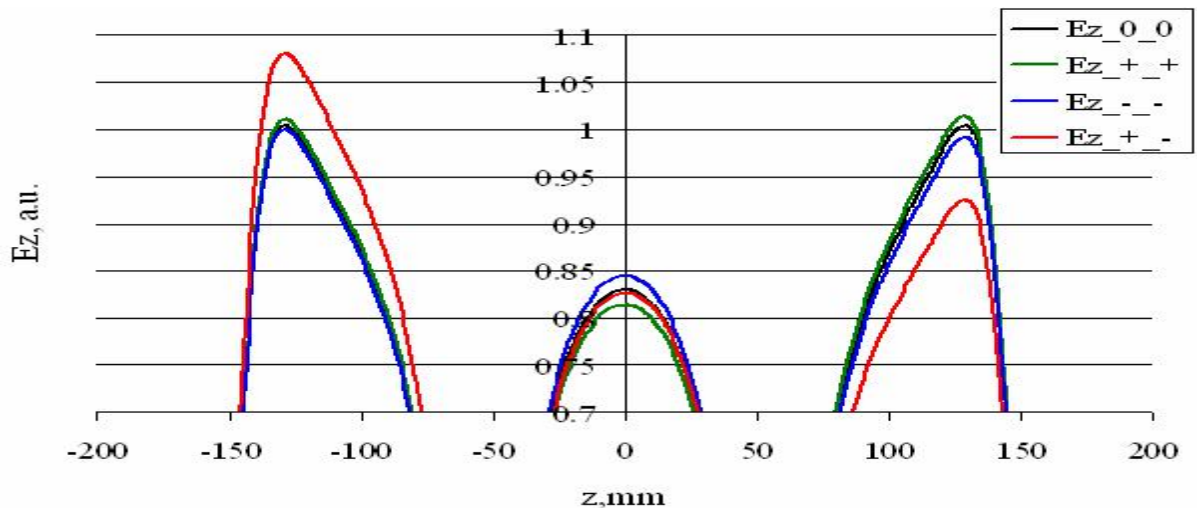

Figure A5. Distributions of |Ez| along cavity axis with positive (+) or negative (-) detuning, 360 kHz in value, in end cells.

Let us to consider detuning of end cells at the frequency $\delta f$=360 kHz. As it is shown in Section 7, see equations (20) and (21), it corresponds to three standard deviations of cell frequencies for mechanical treatment with precision 10 $\mu$m. Distributions of |Ez| along cavity axis for different detuning combinations, positive (+) or negative (-), are illustrated in Fig.

A5. Naturally, cells detuning with opposite signs provides a maximal effect on uniformity of $E_z$ distribution. From plots in Fig. A5 we conclude – RF tuners in end cells are required both to control cells, and cavity frequencies, and to tune required field distribution along axis for operating mode.

**A1.2. RF coupler cell design.**

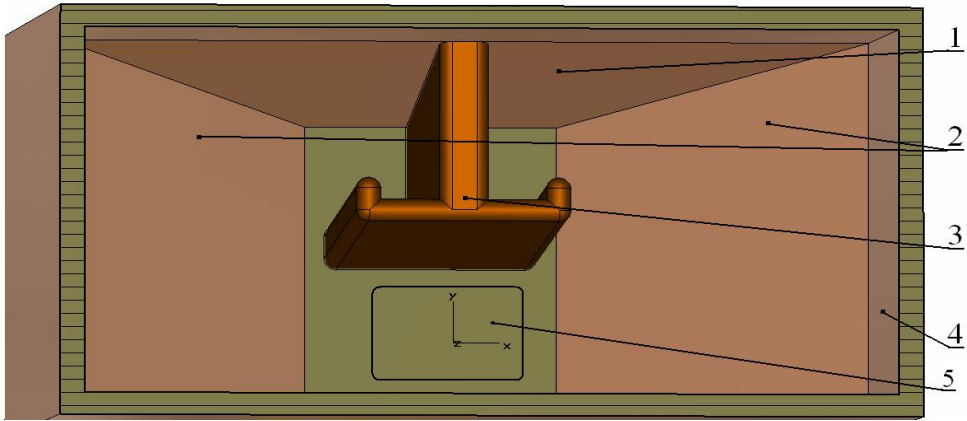

Figure A6. Schematic drawing of waveguide, tapered in both directions, with T-load. 1- non-symmetric tapering in waveguide height, 2- tapering in waveguide width, 3 – T-load, 4- regular waveguide part, 5- slot position.

Design of RF coupler cell all time is the point of special attention. As it is illustrates in Section 6.2, at the matching slot takes place the maximal surface temperature and related stress. In the considered cavity special cone surface, see Fig. A1, Fig. A2, is foreseen and RF coupler introduced in the end cell. It can be shown, [20], [39], that coupling of the cavity with the waveguide depends on magnetic field strength at both sides of matching slot. To increase magnetic field from waveguide side, waveguide tapering in height can be applied, see Section 2.4 and Fig. 7. It results in magnetic field enhancement at 1.6 times at the slot, as compared to regular waveguide. In this design we apply another solution with the special unit, named as T-load. The physical explanation for this unit one can find in [39]. The field enhancement is 2.54, resulting in possibility of shorter slot. Definition of matching slot dimensions was performed in direct CST simulations and results in very short slot (LxW) 26.38mm*29.0 mm. In simulations copper conductivity was reduces to 85% from ideal case, expecting naturally lower real quality factor after construction. Dependencies of S11 elements on frequency in a narrow and wide frequency ranges are shown in Fig. A7. Natural decreasing in coupler cell frequency due to matching slot opening was compensated by reduction of the drift tube length (increasing L2 length for this cell). Until rf coupler cell frequency is unchanged, distribution of fields and RF loss density at the cavity surface is unchanged too, as one can estimate from Fig. A8a. Considering RF density at the surface of waveguide in vicinity of matching slot, Fig. A8b, we do not see significant RF losses and additional cooling for this region is not required.

**A1.3. RF probes design.**

To avoid possible uncertainness in field measurements in the positions of decaying field below cut-off, here we propose direct measurements with RF antenna (assuming Gun 5 antenna, [17]) probe from cavity cell, see Fig. 9a.

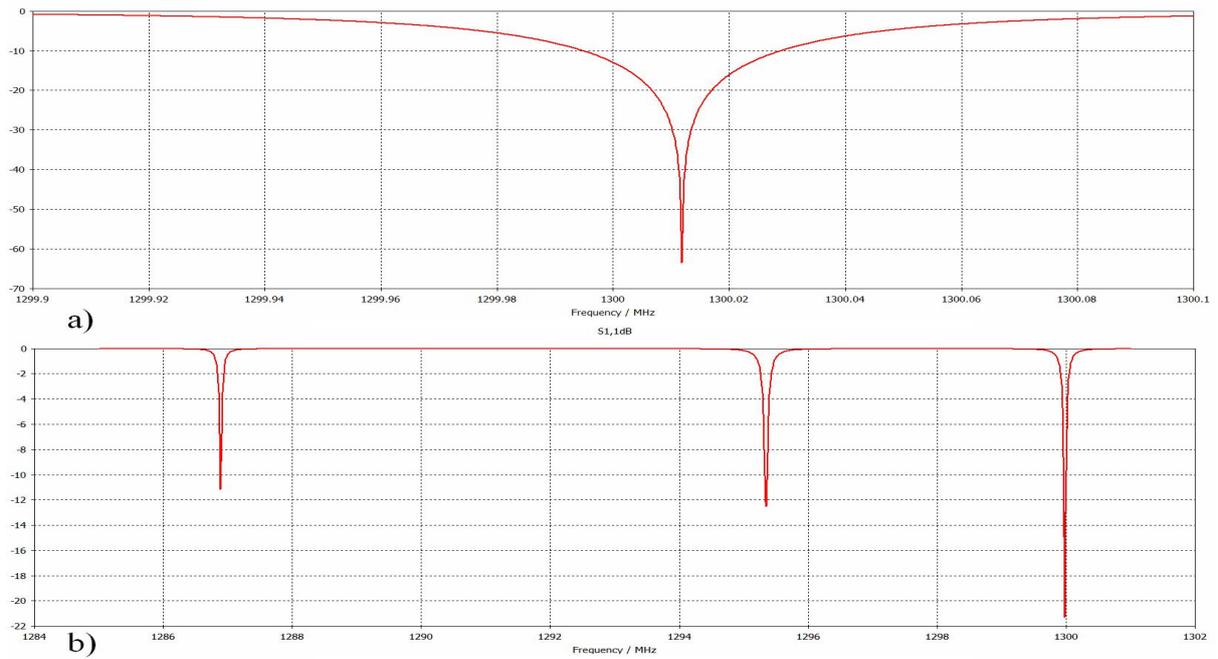

Figure A7. Dependencies of S11 elements on frequency in a narrow, (a), and wide, (b) frequency range.

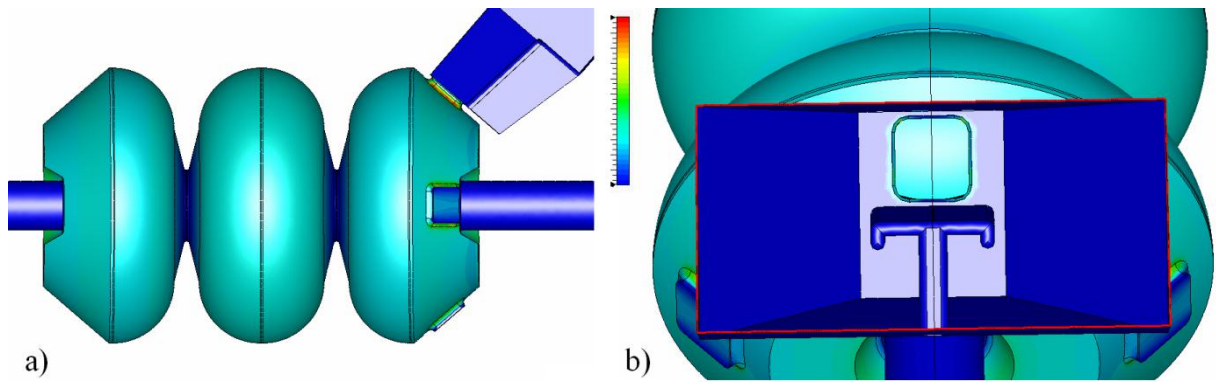

Figure A8. Distributions of RF loss density at the cavity surface with RF coupler, (a), and RF loss density in waveguide in vicinity of matching slot, (b).

At least one probe is required for measurements and a place for probes is foreseen in end cells. But, as it is shown in Session A1.1, field distribution is sensitive to deviations of cells frequencies.

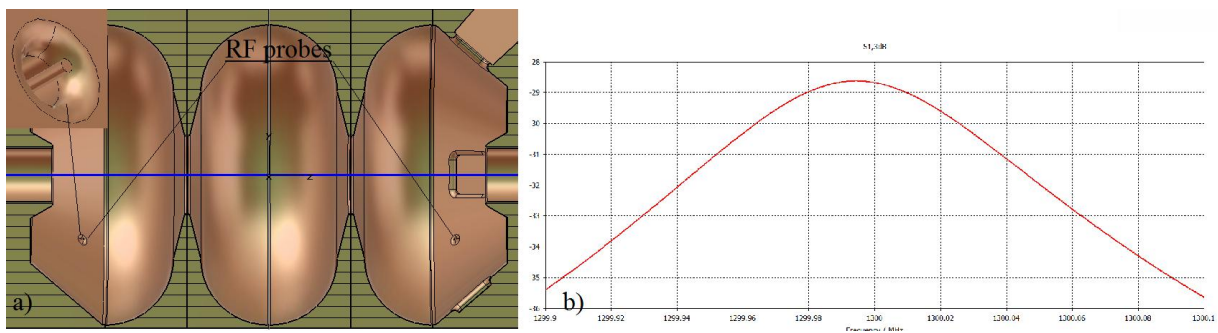

Figure A9. Proposal for RF antenna probes in cavity cells, (a), and S13 dependence for probe tip at the level of cavity surface.

That case the second probe may be useful and probes to be calibrated with bead pull measurements. In Fig. A9b is shown calculated S13 dependence when probe tip is at the level of cell surface. Anyhow, probes should be tuned and required RF signal, ~1W, S13 ~-40 dB, can be obtained with more deep probes position.

### A1.4. Multipacting.

Geometry modifications in the cavity under consideration lead to less uniformity in fields distributions. From physical arguments, it a less comfortable for stable multipacting discharge. Nevertheless, study was performed following to standard procedure, described in Section 5. Data of secondary emission yield corresponding to annealed OFC copper were applied. In Fig. A10 are shown plots for number of electrons vs time for different values of effective voltage at the cavity. Cloud of emitted electrons can survive sufficient time for low voltage value at the cavity. With voltage increasing life time of electron cloud drops fast, as one can conclude from plots in Fig. A10. No conditions for a stable multipacting discharge were found in the vicinity of operating voltage.

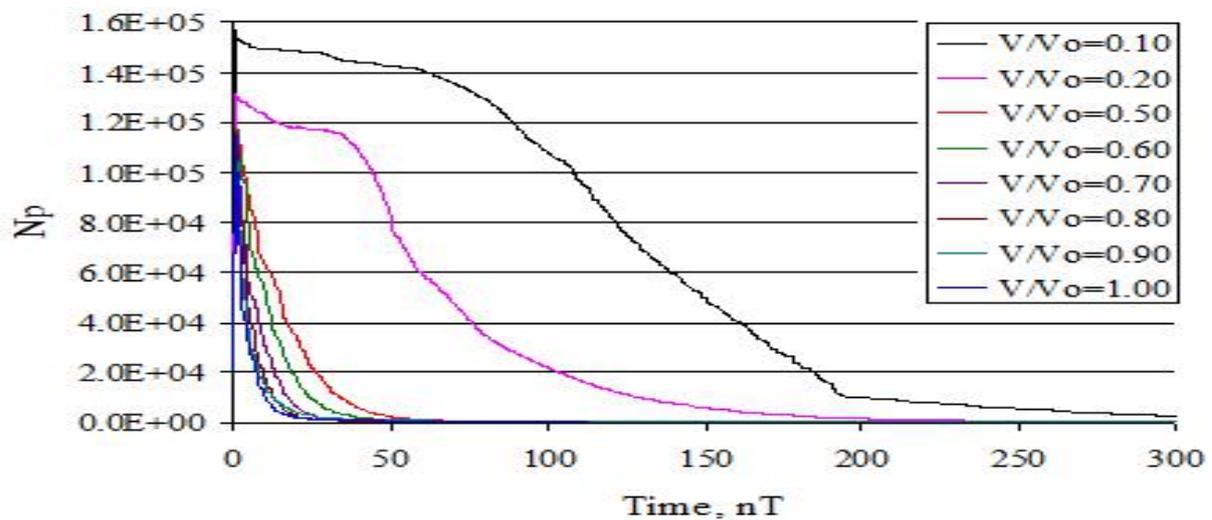

Figure Figure A11. Plots of electrons number vs number of RF periods for different voltage value at the cavity.

### A1.5. Cavity pumping and multipoles compensation.

Cavity can be pumped out trough driving waveguide. Also a special hole for pumping can be foreseen. It depends on additional equipment placing around and should be decided at the stage of technical development. Decision will define also dimensions of blind holes to compensate multipole additions in field distribution. Currently for this cavity this topic is not considered in details. But blind holes can be placed without problems, see Fig. A1. Technique for additions definition is described well in Section 3.1 and definitions of dimensions for compensating holes is the point of ordinary RF simulations.

### A2. Structural analysis

For cavities, operating in CW operating mode the study of coupled thermal effects is an inevitable part of design study.

### A2.1 Cooling circuit.

Design of cooling circuit with internal flow distributions is applied. Ideas and performances of such solution are motivated and described in [17]. The circuit includes small channels for cooling of outer cylindrical surface of the cavity, Fig. A11.a and channels for radial parts cooling, Fig. A11.b. The total circuit is shown in Fig. A11.c. Cooling of conical parts for the cavity with RF coupler, RF tuners and RF probes is assumed with a natural heat conductivity of copper. Technology of cavity construction with such cooling scheme is similar to one for DESY L-band Gun cavities.

All channels are calibrated for the same pressure drop of 35.3 kPa between input/output outlets and for stable flow without swirls, as it is shown in Fig. A12.b in channels for irises cooling. In Fig. A13 are shown distribution of pressure at the surface of channel for outer cavity surface cooling, Fig. A13.a, and velocity distribution in channels cross section, (c).

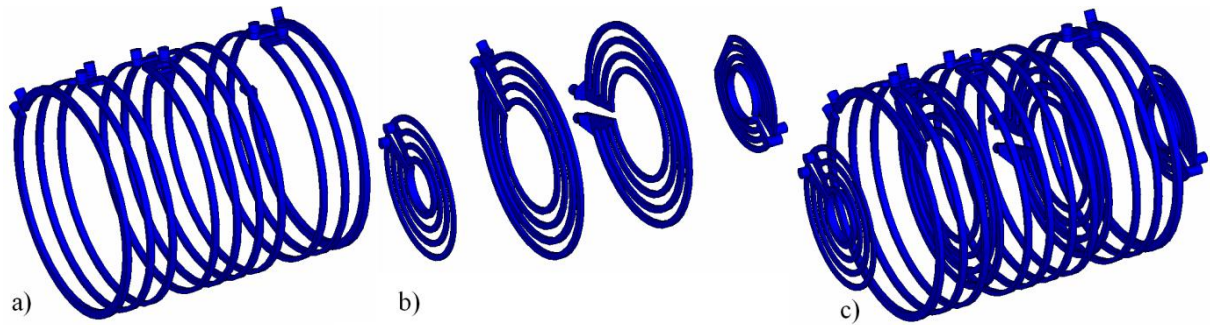

Figure A11. Cooling channels for outer cavity surface, (a), radial parts, (b) and the total circuit, (c).

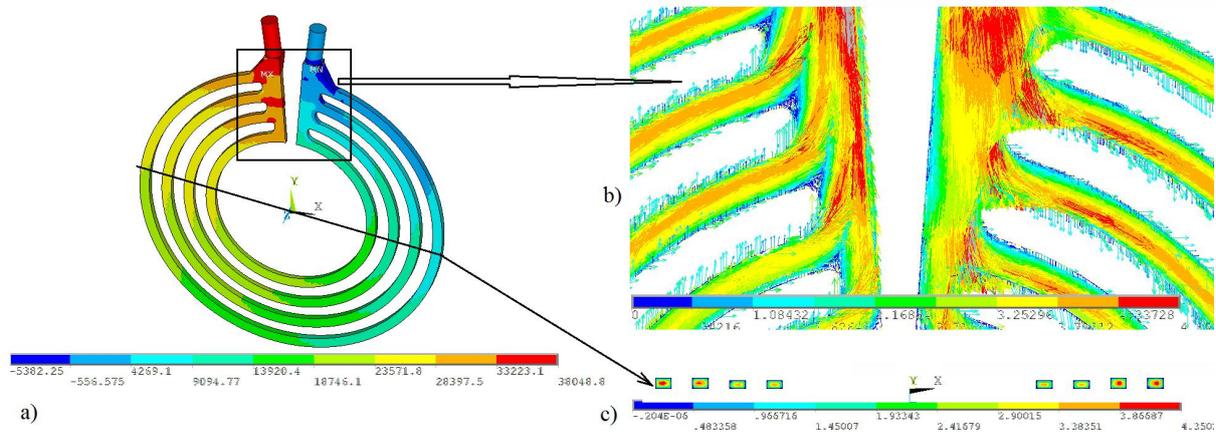

Figure A12. Distribution of pressure at the outer surface of channel for iris cooling, (a) vector plot of flow velocity in the part of flow distribution (b) and velocity distribution in channels cross section, (c).

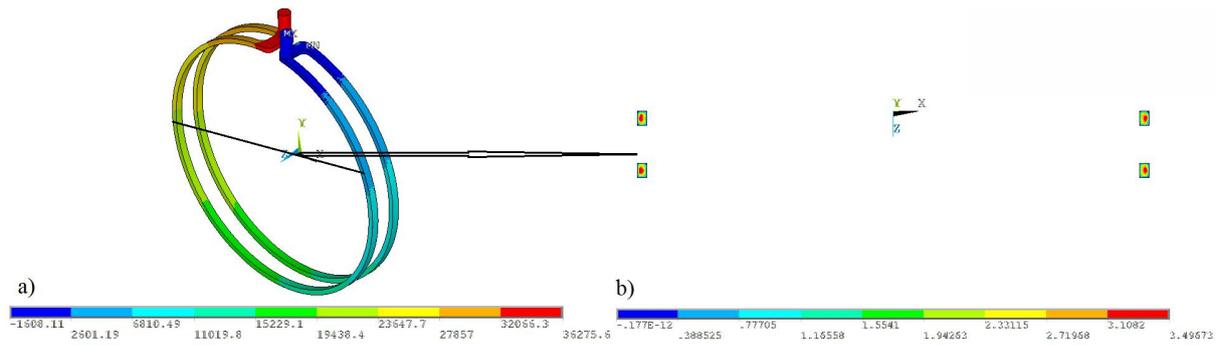

Figure A13. Distribution of pressure at the surface of channel for cavity outer surface cooling, (a) , and velocity distribution in channels cross section, (c).

As one can conclude from Fig. A12.c and Fig. A13.b, the maximal average flow velocity is small channels is of ~2 m/sec. Cooling capability of channels, see Fig. A12.c as example, is fitted with profile of RF losses. The expected water consumption is of 4 m3/h.

## A2.2 Thermal effects.

Simulations of cavity cooling was performed in self-consistent approach, starting from simulations of turbulent flow distributions in cooling channels, see [17] as an example. In simulations increased at 15% value of calculated RF power dissipation, see Table A2, was applied as more close to real value. Calculated distribution of temperature at the outer surface of the cavity is shown in Fig. A14.a and the maximal temperature rise near matching slot is of ~ 18.1 C. In the cavity cross section, with a new scale, distribution of temperature is shown in Fig. A14.b.

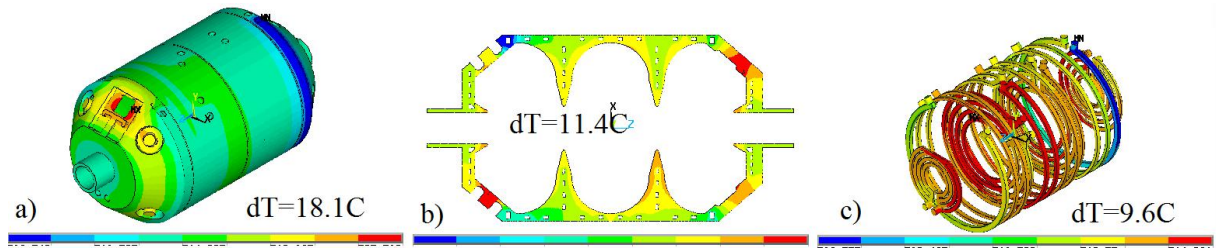

Figure A14. Temperature distributions at the cavity surface, (a), in cavity cross section, (b) and at the surface of cooling channels, (c).

Distribution of displacements at the surface of total cavity is shown in Fig. A15.a and corresponding stress distribution in cavity copper part is illustrated in Fig. A15b. The maximal stress value of 25.1 MPa at the surface of matching window is in the safe limits.

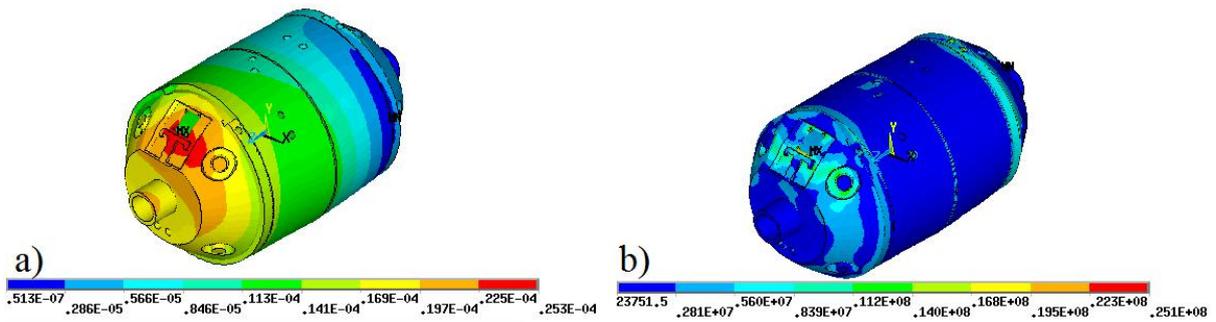

Figure A15. Distributions of displacements at the cavity surface, (a) and stress distribution in copper cavity body, (b).

### A2.2 Thermal induced frequency shift.

Calculated value of induced frequency shift of -827 Hz is very small for cavities in CW mode. It was not a design goal and is unexpected. This phenomenon was multiply checked for different water flow and in different approaches for cooling simulations – engineering, (as described in Section 6.1) and self-consistent. All times very small, several kHz, frequency shift. Mostly reasonable explanation is in mutual compensation of electric part and magnetic part, see (22) for this cavity design.

This small frequency shift is much less than expected width of resonant curve of the cavity with the matched RF coupler ~100 kHz and tuned cavity can start into operation without intermediate operations.

### A2.3 Effect of RF tuners.

As one can see from Fig. A1, sufficient space is foreseen to place RF tuner with sufficient diameter of blind holes. In Fig. A16. For diameter of blind hole 32 mm shift of one tuner at 1 mm results if cavity frequency shift of 95 kHz. So, eight tuner will provide tuning range +- 760 kHz, which is sufficient to compensate at least frequencies deviations due to dimensions deviations.

This small frequency shift is much less expected width of resonant curve of the cavity with the matched RF coupler ~100 kHz and tuned cavity can start into operation without intermediate operations.

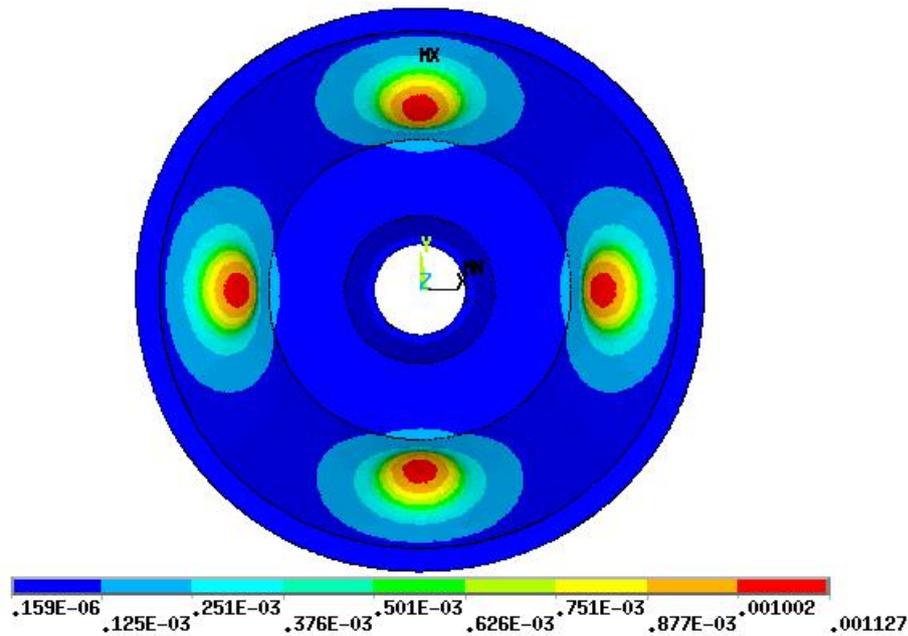

Figure A16. Distributions of internal surface displacements by RF tuners.

**A3. Summary.**

An additional design of CW buncher cavity is presented. This design strongly based on mastered experience of DESY L-band cavities development and construction. Stainless steel jacket serves as the strong support for mounting input/output outlets of cooling circuit and cavity mounting at the girder, providing rigid mechanical design. Cavity is optimized for slightly higher shunt impedance and improved separation in frequency with nearest mode. The space for effective RF tuners is foreseen in end cells. Gun 5 style RF antenna probes are suggested for direct measurements from cavity cells. Effective cooling circuit is proposed assuming the same pressure drop for all cooling channels and resulting in a lower cavity temperature rise and lower water consumption. The maximal stress value is below yield stress for annealed OFE copper with reliable reserve. Very small frequency shift, induced by thermal deformations results in simplified start of operation.